\documentclass[a4paper,oneside,final,notitlepage,onecolumn,12pt,english]{article}

\usepackage{ifpdf,epsfig,array,amsmath,amssymb,psfrag,graphicx}
\usepackage{ulem} % \xout{removed}
\ifpdf
\usepackage[colorlinks=true,filecolor=blue,linkcolor=blue,citecolor=blue,urlcolor=blue]{hyperref}
\usepackage[usenames,dvipsnames]{color}
\else
\usepackage[usenames,dvips]{color}
\fi
\definecolor{blue}{rgb}{0,0.2,0.8}
\definecolor{red}{rgb}{1,0,0}

\DeclareFontFamily{OT1}{rsfs}{} \DeclareFontShape{OT1}{rsfs}{m}{n}{
<-7> rsfs5 <7-10> rsfs7 <10-> rsfs10}{}
\DeclareMathAlphabet{\mycal}{OT1}{rsfs}{m}{n}

\def\scri{{\mycal I}}%skraj
\def\scrip{\scri^{+}}%
\def\scrim{\scri^{-}}%

\DeclareFontFamily{OT1}{rsfs}{} \DeclareFontShape{OT1}{rsfs}{m}{n}{
<-7> rsfs5 <7-10> rsfs7 <10-> rsfs10}{}
\DeclareMathAlphabet{\mathscr}{OT1}{rsfs}{m}{n}

\psfrag{I+}{$\scrip$}
\psfrag{I-}{$\scrim$}
\psfrag{H+}{$\mathcal H_+$}
\psfrag{H-}{$\mathcal H_-$}
\psfrag{i+}{$i^+$}
\psfrag{i-}{$i^-$}
\psfrag{i0}{$i^0$}
\psfrag{I}{I}
\psfrag{II}{II}
\psfrag{III}{III}
\psfrag{IV}{IV}

\newcommand{\beq}{\begin{equation}}
\newcommand{\eeq}{\end{equation}}
\newcommand{\bea}{\begin{eqnarray}}
\newcommand{\eea}{\end{eqnarray}}

\newcommand{\intd}{\mathrm{d}}

\newcommand{\ii}{\mathrm{i}}

\newcommand{\CQG}{{\it Class. Quant. Grav. }}
\newcommand{\PRD}{{\it Phys. Rev.} D }

\begin{document}

\title{Numerical investigation of the late-time Kerr tails}
\author{\small 
Istv\'{a}n R\'{a}cz\thanks{email: iracz@rmki.kfki.hu} \ and  G\'{a}bor 
Zs.\,T\'{o}th\thanks{email: tgzs@rmki.kfki.hu}
\\ %EndAName
\small RMKI, \\
\small  H-1121 Budapest, Konkoly Thege Mikl\'os \'ut 29-33.\\
\small Hungary
}
\maketitle

\begin{abstract}
The late-time behavior of a scalar field on fixed Kerr background is examined in a
numerical  framework incorporating the techniques of conformal compactification and 
hyperbolic initial value formulation. 
The applied code is 1+(1+2) as it is based on the use of the spectral method in the angular
directions while in the time-radial section fourth order finite differencing, along
with the method of lines, is applied. 
The evolution of various types of stationary and 
non-stationary pure multipole initial states are investigated. The asymptotic decay rates are determined not only 
in the domain of outer communication but along the event 
horizon and at future null infinity as well. The decay rates are found to be different for  
stationary and non-stationary initial data, and they also  
depend on the fall off properties of the initial data 
toward future null infinity. 
The energy and angular momentum transfers are found 
to show significantly different behavior in the initial phase of the time evolution.
The quasinormal ringing phase and the tail phase are also investigated. In the tail phase, 
the decay exponents for the energy and angular momentum losses at $\scrip$ are found to be smaller than at the horizon 
which is in accordance with the behavior of the field itself and 
it means that at late times the  energy and angular momentum 
falling into the black hole become negligible in comparison with the  energy and angular momentum 
radiated toward $\scrip$.  
The energy and angular momentum balances are used as additional verifications of 
the reliability of our numerical method.
\end{abstract}

\vfill\eject

\section{Introduction}
%\label{}

The late-time behavior of linear fields propagating on Schwarzschild background 
has been studied in much detail since the pioneering work of Price \cite{Price}. 
By now it is well understood and there is also compelling numerical 
evidence supporting the theoretical picture. The analysis is considerably 
simplified by the fact that, due to spherical symmetry of Schwarzschild spacetime, 
the perturbing field can be uniquely represented as a linear combination of 
spherical harmonics and the coefficients of the separate multipole components 
evolve independently. It has been justified by several authors that for 
compactly supported, generic, i.e.\ non-stationary
initial data, an $(l,m)$ multipole component decays at late times 
as $t^{-(2l+3)}$, regardless of the type---scalar, electromagnetic, 
gravitational, etc---of the perturbing field.

\medskip

Since the Kerr spacetime is not spherically symmetric, the late-time decay on 
Kerr background is more complicated and the 
numerical results had also remained controversial until quite recently. 
Due to the lack of spherical symmetry, even if initially only a single 
multipole component is excited,  inevitably additional multipole components are also 
generated during the time evolution. Fortunately, the axial symmetry along with the 
equatorial reflection symmetry of the Kerr spacetime 
restricts the allowed 
multipole components. Nevertheless, somewhat 
surprisingly, it was found that 
in certain coordinate systems, including the Boyer-Lindquist and ingoing Kerr 
ones, the decay rates of the excited multipole components show a dependence on 
the multipole indices of the initially excited mode that cannot be 
accounted for simply by referring to
the symmetries of the background. 
For scalar perturbations, in the case of non-stationary initial data specifications, 
this sort of behavior was predicted by the analytic investigations of 
Barack and Ori \cite{BO,Barack}, 
and of Hod \cite{Hod1}. In addition, predictions for the decay exponents---relevant for the case with Boyer-Lindquist coordinates---were also  derived in \cite{BO,Barack,Hod1}.
Numerical justifications of these analytic results were also provided by 
Burko and Khanna \cite{BK,BK3}, Tiglio, Kidder and Teukolsky \cite{TKT}, 
Gleiser, Price and Pullin \cite{GPP}, Krivan, Laguna and Papadopoulos \cite{KLP}, 
and  Krivan \cite{Krivan}.

\medskip

To understand the main difficulties showing up in the study of the evolution of a linear field, $\Phi$, satisfying the homogeneous wave equation  
\beq\label{fe}
\nabla_\mu\nabla^\mu\Phi=0\,,
\eeq
on Kerr black hole background, note that while due to spherical symmetry of the Schwarzschild spacetime the family of two-spheres invariant under the action of the rotation group provides a geometrically distinguished and thereby unique foliation of the spacetime, such a unique foliation does not exist in the Kerr case. Although some of the most frequently used coordinates, such as the Boyer-Lindquist or Kerr-Schild coordinates, 
offer immediate choices, the pertinent topological two-spheres do not coincide and, in turn, a pure multipole state with respect to one of these foliations is not a pure one with respect to the other, and vice versa. For a related discussion see e.g.\ \cite{Poisson}, where the explicit foliation dependence of the decay rates was also clearly demonstrated. 
Tiglio, Kidder and Teukolsky \cite{TKT} provided numerical 
confirmation of these analytic indications by comparing evolutions based on Boyer-Lindquist or Kerr-Schild coordinates.
Numerical investigations using Kerr-Schild coordinates were also done by Scheel et al.\ \cite{Scheel}, 
with results that are in agreement with those of \cite{TKT}. 
It was found, in particular, that in Kerr-Schild coordinates the late-time decay rates relevant for the lowest admitted multipole components 
do depend on the multipole indices of the exciting pure initial data only up to the extent that can be deduced from   
the axial and reflection symmetries of the Kerr spacetime. Note, 
however, that even in this simple case the decay rates of the higher multipole components differ from the corresponding decay rates in the Schwarzschild case. 
In addition to the non-unique choice of the two-spheres, 
the consequences of using  different time slicings and the associated loss of the generality of the initial data specifications also increase the complexity of the picture. 
A significant contribution to the clear up is due to Tiglio, Kidder and Teukolsky \cite{TKT} who carried out a systematic comparison of the decay rates based on the use of Boyer-Lindquist and Kerr-Schild coordinates.

\medskip

In favoring the non-Kerr-Schild type coordinate systems, Burko and Khanna \cite{BK3} provided numerical justification of 
the fact that the decay rates of tails in Boyer-Lindquist and in ingoing Kerr coordinates agree,  
in spite of the fact that the time slices are different as the latter are horizon penetrating ones.  
They also claimed that these coordinate systems belong, in a certain sense, 
to an equivalence class in which pure multipole initial data are distinguished 
and their indices are uniquely determined by the late time decay rates yielded by them.
By making use of the relation between the Boyer-Lindquist and ingoing Kerr coordinates---see equations (\ref{tr1}) - (\ref{tr2})---they argued 
in  \cite{BK3} that a spherical harmonic mode with indices $(l,m)$ in Boyer-Lindquist coordinates is also a spherical harmonic mode 
with the same values for $(l,m)$ in ingoing Kerr coordinates, which may explain the similar behavior of the tails in these two coordinate systems. 
The latter point, however, needs further justification and more rigorous arguments. 

\medskip

Providing a verification of the proposal of Burko and Khanna \cite{BK,BK3}, in a slightly different setup, 
Zenginoglu and Tiglio \cite{ZT} carried out a numerical investigation of the decay rates based on the use of
conformal compactification, such that the compactified versions of both the Boyer-Lindquist and the ingoing Kerr coordinates were applied. 
Although their results for the asymptotic decay rates are not very
extensive, the reported values are in good agreement with that of \cite{Hod1,BK,BK3}. In particular, they support the suggestions of 
Burko and Khanna \cite{BK3}. One of the main advantages of using conformal compactification is that 
the decay rates at 
future null infinity, $\scrip$, may also be determined. The pertinent decay rates reported in \cite{ZT} are in agreement with the predictions of \cite{Hod1}.

\medskip

The aim of this paper is to strengthen the developing consensus by presenting the results of our numerical investigations on the late-time evolution of 
tails in the Kerr spacetime. The applied analytic setup is based on the use of the techniques of conformal compactification and hyperboloidal 
initial value formulation. To simplify the presentation and also to provide a straightforward way to derive the basic equation 
to be solved---instead of a technical but covariant argument---we have chosen a non-covariant one in which two subsequent coordinate 
transformations are applied. In addition to the usually chosen quickly decaying initial data specifications some slowly decaying cases, 
which have not been studied yet, with $1/r^k$ fall off for $\Phi$ toward $\scrip$  are also considered. Similarly, to get some more insight 
about the time evolution of the scalar field on Kerr background, not only the axially symmetric case but in a few cases excitations with azimuthal 
indices $m=\pm 1$ are also studied. In addition to the usual convergence criteria the energy and angular momentum conservation laws are also monitored. 
Our results concerning the associated energy and angular momentum transports also provide some important clue about the evolution of linear fields 
on rotating black hole backgrounds.

\medskip

The paper is organized as follows. In Section \ref{sec.2}, the applied analytic and numerical 
framework is introduced. In
Section \ref{sec.3}, our main results are presented, and Section \ref{sec.4} contains our final remarks. Throughout this paper, geometric units, 
with $c=G=1$, are applied.

\section{Preliminaries}
\label{sec.2}

This section is to recall some of the basics of the applied mathematical and numerical framework.

\subsection{The analytic setup}

The Kerr metric in Boyer-Lindquist coordinates $(t,r,\theta,\phi)$ reads as
\beq
ds^2=-\left(1-\frac{2Mr}{\Sigma}\right)dt^2-\frac{4 a r M\sin^2\theta}{\Sigma}dt d\phi
+\frac{\Sigma}{\Delta}dr^2 + \Sigma d\theta^2 +\frac{\Gamma }{\Sigma}\sin^2\theta d\phi^2,
\eeq 
where
$\Sigma=r^2+a^2\cos^2\theta$, $\Delta=r^2-2Mr+a^2$, $\Gamma=(r^2+a^2)^2-a^2 \Delta\sin^2\theta$. $M$ and $a$ 
denote the mass and the angular momentum per unit mass of the Kerr black hole, respectively.

The desired new coordinates, which have the advantages that the 
time slices are horizon penetrating and
connect to future null infinity  
such that no boundary conditions are needed,
are introduced by performing the following two subsequent coordinate transformations. 

\begin{figure}[htbp]%[htbp!]
\vskip-3.5cm\begin{center}
\ {} \quad\quad\quad\quad\quad\quad\quad\quad{\quad\quad\quad\quad\quad\quad\quad\quad{\hskip4cm\includegraphics[width=130mm,angle=270]{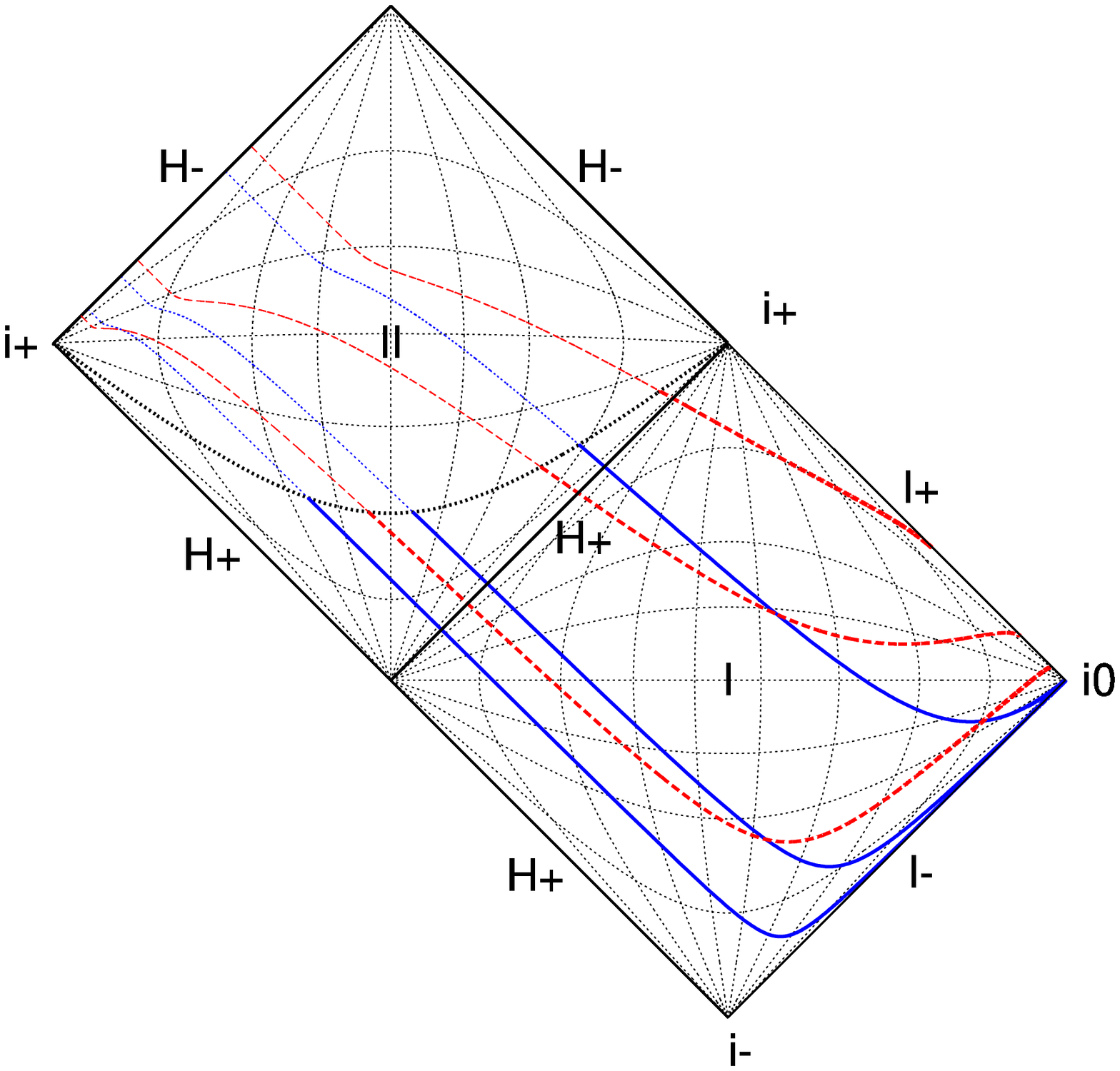}}}\\ \vskip-3.5cm{\hskip-9cm \includegraphics[width=90mm,angle=270]{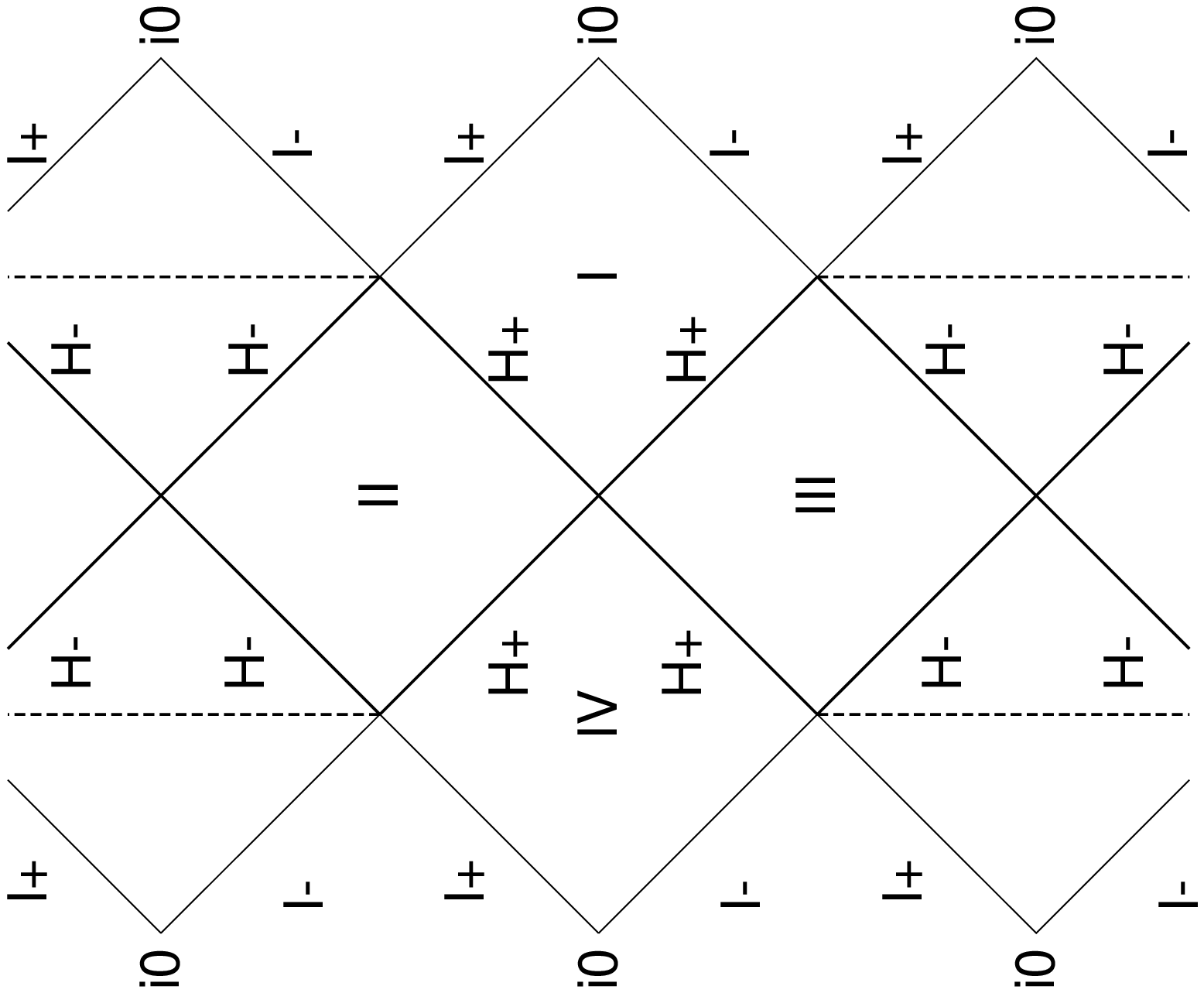}}
\end{center}
\caption{\label{fig1}
At the left lower block a schematic version of the Carter-Penrose diagram of maximal analytic extension of the Kerr spacetime is shown while on the top a 
part of this diagram is enlarged to show in more detail regions I and II. In these regions, the Boyer-Lindquist $t$ and $r$ coordinate lines are indicated 
by the thin dotted lines. Some of the $\tau=const$ and $T=const$ time-slices are indicated, for the case with $M=1$ and $a=0.5$, by the solid blue and 
dashed red lines, respectively. 
The red lines illustrate that the $(T,R)$ coordinates are horizon penetrating and connect to $\scrip$. 
It is worth mentioning that constant time-slices of the Kerr-Schild coordinates coincide with that of the ingoing Kerr coordinates and these two 
coordinate systems differ only in the applied foliation by topological two-spheres.
}
\end{figure}

\begin{itemize}

\item[(1)] First, the time and azimuthal coordinates, $t$ and $\phi$, of the Boyer-Lindquist system are replaced by $\tau$ and $\varphi$ defined as 
\begin{eqnarray}
\label{tr1}
%\tau & = & t-r+r^* \\
\tau & = & t-r+\int\intd r\, \frac{r^2+a^2}{\Delta} \\
\label{tr2}
%\varphi & = & \phi+ \phi^* , 
\varphi & = & \phi+ \int\intd r\, \frac{a}{\Delta} \,.
%\end{eqnarray} 
%where the functions $r^*$ and $\phi^*$ are given as 
%\begin{eqnarray}
%\label{tr3}
%r^* & = & \int\intd r\, \frac{r^2+a^2}{\Delta} \\
%\label{tr4}
%\phi^* & = & \int\intd r\, \frac{a}{\Delta}\,.
\end{eqnarray}
The coordinates $(\tau,r,\theta,\varphi)$ are usually referred to as ingoing Kerr coordinates. 
As opposed to the Boyer-Lindquist coordinates, these coordinates remain regular at the outer horizon $\mathcal{H}_+$ separating the domain of 
outer communication from the interior domain between the inner and outer event horizon. These regions are indicated by domains I and II in 
Figure \ref{fig1}, respectively. As the radial coordinate, $r$, is not changed, the inner and outer event horizons, $\mathcal{H}_-$ and $\mathcal{H}_+$, 
are located at $r=r_\mp=M\mp\sqrt{M^2-a^2}$, respectively.  As indicated in Figure \ref{fig1}, the $\tau=const$ time-slices 
intersect $\mathcal{H}_+$, justifying the terms `ingoing Kerr coordinates' and `horizon-penetrating slicing'. Note also that in the $r\rightarrow \infty$ 
limit the $\tau=const$ time-slices tend to spacelike infinity, denoted by $i^0$.

\item[(2)] Second, we replace the ingoing Kerr coordinates $\tau$ and $r$ by the new time coordinate $T$ and by the 
compactified radial coordinate $R$, defined via the implicit relations
\begin{eqnarray}
\label{tr5}
\tau & = & T+\frac{1+R^2}{1-R^2}-4M\log(|1-R^2|)\\
\label{tr6}
r & = & \frac{2R}{1-R^2}\,.
\end{eqnarray}
The transformation (\ref{tr5}) -- (\ref{tr6}) yielding the new 
coordinates $(T,R,\theta,\varphi)$ differs from the conformal transformation proposed by Moncrief \cite{Moncrief} 
(see also \cite{FR1,FR2}) only in the logarithmic term in (\ref{tr5}) which---according to the conclusions of \cite{Anil}---is inevitably present 
in most of the analogous conformal scri fixing compactifications of the Kerr spacetime.
The $T=const$ surfaces (see Figure \ref{fig1} for an illustration), 
tend to future null infinity ($\scrip$) as $r\rightarrow \infty$, which is represented by the $R=1$ boundary in the new 
coordinates\footnote{Note that some of the coordinate components of the metric tensor blow up at $R=1$, 
which in the framework of conformal compactification is usually compensated by replacing the physical metric $g_{\mu\nu}$ by a 
non-physical one $\tilde g_{\mu\nu}=\Omega^2 g_{\mu\nu}$, where $\Omega$ is the conformal factor. 
In the particular case considered here, $\Omega=\frac{1-R^2}{2R}$.} 
and are spacelike in a neighborhood of the domain of outer communications.    
It is straightforward to see that the coordinate basis fields $({\partial}/{\partial T})^\mu$ and $({\partial}/{\partial \varphi})^\mu$ are Killing vectors.
A major advantage of these coordinates is that one can place the inner and outer boundaries of the domain of numerical investigations inside 
the black hole region and to future null infinity. Since $\mathcal{H}_+$ and $\scrip$ are null hypersurfaces admitting
the waves to leave the domain of outer communication but not allowing any to enter, such a choice considerably reduces spurious boundary effects 
in the domain of outer communication (see also \cite{DR1} for a discussion on the effects of boundary conditions). Furthermore,  
there is no need to impose any  sort of sophisticated boundary condition  and we find it to be completely satisfactory to require simply 
the field equation to hold at the edges. 
In our numerical calculations, the inner boundary (see the thick dotted line in Figure \ref{fig1}) is  placed slightly beyond the event horizon 
and the outer boundary precisely at $R=1$.
\end{itemize}

In addition to the above transformations, we also introduce the rescaled field variable  
\beq
\label{eq.resc}
\Psi=\frac{2 R}{1-R^2}\,\Phi.
\eeq
The pertinent form of (\ref{fe}) for $\Psi$ is regular in both of the spacetime regions $\rm I$ and $\rm II$, 
it remains regular\footnote{It is known 
that a conformal compactification, 
which is achieved in the present case by the introduction of the coordinates $(T,R,\theta,\varphi)$, 
also needs to be compensated by a rescaling of the type (\ref{eq.resc}) in the field variable, 
otherwise the relevant form of (\ref{fe}) for $\Phi$, exactly as the physical metric does, becomes singular at $\scrip$. (See also \cite{Zenginoglu} 
for an interesting related argument applicable in investigations carried out in the frequency domain.)} up to and including 
the boundary at $R=1$, and it can be given in the form 
\begin{eqnarray}
\label{eq.de}
&&\hskip-.5cm\partial_{TT}\,\Psi = \frac{1}{a_{TT}^{(0)}+a_{TT}^{(2)}\,Y_2^0}(a_{RR}\,\partial_{RR}+a_{TR}\,\partial_{TR} 
+a_{T\varphi}\,\partial_{T\varphi}+a_{R\varphi}\,\partial_{R\varphi} \\ 
&&\hskip-.5cm\phantom{\partial_{TT}\,\Psi = \frac{1}{a_{TT0}+a_{TT2}\,Y_2^0(\theta,\varphi)}(aa_{T}\,\partial_{T} }
+a_{T}\,\partial_{T}+a_{R}\,\partial_{R}+a_{\varphi}\,\partial_{\varphi}
+a_0 +a_{\Delta}\Delta_{S^2})\,\Psi\,, \nonumber
\end{eqnarray}
where $Y_2^0=Y_2^0(\theta,\varphi)=\frac{1}{4}\sqrt{\frac{5}{\pi}}(3\cos^2\theta-1)$ is the spherical harmonic function 
with $l=2$, $m=0$ and $\Delta_{S^2}$ stands for the spherical Laplace operator
\beq
\Delta_{S^2}=\frac{1}{\sin\theta}\,\frac{\partial}{\partial \theta}\,\left(\sin\theta\frac{\partial}{\partial\theta}\right) + \frac{1}{\sin^2\theta}\, \frac{\partial^2}{\partial \varphi^2}\,.
\eeq 
It is an important property of (\ref{eq.de}) that all the coefficients 
$a_{RR}$, $a_{TR}$, $a_{T\varphi}$, $a_{R\varphi}$, $a_{T}$, $a_{R}$, $a_{\varphi}$, $a_{0}$, $a_{\Delta}$, $a_{TT}^{(0)}$ and $a_{TT}^{(2)}$ 
are functions
of $R$ exclusively, i.e.\ they do not depend on $T$, $\theta$ and $\varphi$. 
Explicit expressions for 
these coefficients are given in Appendix A.
From equation (D.14) of \cite{Wald} and from the vanishing of the scalar curvature of the Kerr metric, it follows that 
$g^{\mu\nu}\nabla_\mu\nabla_\nu \Phi= \Omega^{3} (\tilde{g}^{\mu\nu}\tilde{\nabla}_\mu\tilde{\nabla}_\nu -\frac{1}{6}\tilde{S})  \Psi$, where 
$\tilde{\nabla}_\mu$ is the covariant derivative corresponding to the rescaled metric $\tilde{g}_{\mu\nu}=\Omega^{2} g_{\mu\nu}$,
$\tilde{S}$ is the scalar curvature of $\tilde{g}_{\mu\nu}$,
$\tilde{g}^{\mu\nu}=\Omega^{-2} g^{\mu\nu}$
and $\Omega^{-1}=\frac{2 R}{1-R^2}$. Using this equation,  the relation
$a_0=\frac{\Omega^{2}}{6 g^{TT}}\tilde{S}(a_{TT}^{(0)}+a_{TT}^{(2)}\,Y_2^0)$ can be obtained. Note that while $a_0$ depends only on $R$, 
$\tilde{S}$ also depends on $\theta$.
The reason for writing the field equation in the form (\ref{eq.de}), where it is solved for $\partial_{TT} \Psi$,  
is that it is this form that is needed for the application of the Runge-Kutta method, which is the method that is used
to compute the time evolution of the field.
It is a further important property of (\ref{eq.de}) that  
the denominator on the right hand side is non-vanishing in a neighborhood of the domain of outer communication.

\subsection{The numerical method}

In solving (\ref{eq.de}), the angular dependence of the field is handled by spectral decomposition based on the foliation by two-spheres given by 
the $t,r=const$---or equivalently, $T,R=const$---surfaces. Accordingly, $\Psi$ is replaced by the series of spherical harmonics
\beq
\label{ser}
\Psi=\sum_{l,m} \psi_l^m(R,T)\,Y_l^m(\theta,\varphi)\,,
\eeq
which reduces the evolution problem to determining that of the coefficients $\psi_l^m(R,T)$ in the $T - R$ section. 
In practice, the series (\ref{ser}) has to be truncated to a finite sum; nevertheless, 
as it converges quickly, the desired precision can always be guaranteed by keeping sufficiently many terms. Note that (\ref{eq.de}) has a relatively 
simple structure, in which only the denominator, in particular its second term, 
is responsible for a mixing between different spherical harmonic modes.
Following the basic ideas described in more detail in \cite{alir,gridripper}, the division by the term $a_{TT}^{(0)}+a_{TT}^{(2)}\,Y_2^0$ is converted into 
a multiplication by applying the identity 
\beq
\label{nser}
1/(1+x)=1-x+x^2-x^3+x^4-x^5\cdots\,,
\eeq
along with the choice $x=a_{TT}^{(2)}\,Y_2^0/a_{TT}^{(0)}$ and the use of the relation 
\beq
\label{cgc}
Y_{l_1}^{m_1} Y_{l_2}^{m_2} = \sum_{l_3,m_3} G_{l_1 l_2 l_3}^{m_1 m_2 m_3} Y_{l_3}^{m_3}\,.
\eeq
On the right hand side of (\ref{nser}), one has to keep as many terms as needed to get the error induced by this approximation to be smaller than a 
fixed tolerance. 
In our simulations the error tolerance is chosen to be $10^{-20}$, corresponding to the extended precision arithmetic 
that we use. For $a=0.5$, which is the value that is taken in most of our calculations, 
$11$ terms are kept in (\ref{nser}), which is sufficient since  
$|a_{TT}^{(2)}\,Y_2^0/a_{TT}^{(0)}|< 0.00775$ holds for $a=0.5$. 
The coefficients $G_{l_1 l_2 l_3}^{m_1 m_2 m_3}$, appearing in (\ref{cgc}), 
are closely related to Clebsch-Gordan coefficients and are described in standard mathematical references. 
In addition to the above replacements the well-known relations
\beq
\Delta_{S^2}Y_l^m=-l(l+1)Y_l^m\ \ \ {\rm and}\ \ \ \partial_\varphi Y_l^m=\ii mY_l^m
\eeq
are also applied. Due to the axial symmetry of the Kerr background and the linearity of (\ref{eq.de}), 
the time evolution of the coefficients $\psi_l^m(R,T)$ with different azimuthal indices decouple, which, in particular, 
allows one to investigate their evolution separately. As the background also possesses the equatorial reflection symmetry, 
the coefficients $\psi_l^m(R,T)$ with odd differences in their $l$ indices also decouple, which considerably simplifies 
the study of the evolution of pure multipole initial data.  
In our simulations pure multipole initial data with $l'\le 5$ are considered, where $l'$ is used to denote 
the polar index of the initial data. It is found that 
with such initial data (the complete specification of our initial data can be found in Section \ref{sec.init})   
it is  sufficient
for the study of several modes up to $l=4$ 
to keep only the first $8$ non-zero terms in the multipole expansion (\ref{ser}) of $\Psi$,
due to the fast convergence of this series.
This means that the multipole components 
taken into account are those for which $l=|m|, \dots, |m|+((l'-|m|) \mod 2) +14$, since $l\ge |m|$ for any spherical harmonic function and the non-zero 
modes are those for which $l'-l$ is even.

As the coefficient $a^{(2)}_{TT}$ is proportional to the angular momentum per unit mass  parameter of the Kerr spacetime, it follows immediately from the 
above discussion that the spherical modes completely decouple in the Schwarzschild limit. It is also worth keeping in mind that, 
in virtue of (\ref{tr1}) -- (\ref{tr6}), a spherical harmonic mode with indices $(l,m)$ defined with respect of the Boyer-Lindquist coordinates 
is also a spherical harmonic mode, with the same $(l,m)$ indices, with respect to the compactified ingoing Kerr coordinates $(T,R,\theta,\varphi)$, 
and vice versa. 

\medskip

In solving (\ref{eq.de}), it is recast into a first order form by introducing 
the new auxiliary variables $\Psi_T=\partial_T\Psi$ and $\Psi_R=\partial_R\Psi$, 
and in the $T - R$ section 
the method of lines is applied to compute the time evolution of the multipole coefficients of $\Psi$, $\Psi_T$ and $\Psi_R$.
Fourth order finite differencing is used in the radial direction while for the time evolution
a fourth order Runge-Kutta scheme is applied.  
In order to suppress high frequency instabilities, a standard fifth order Kreiss-Oliger
dissipation term, as proposed in \cite{Gustetal}, is added
to the right hand side of the equations, and the constraint 
$\Psi_R=\partial_R\Psi$ is imposed after each time step.
It is found that $2048$ grid cells are sufficient
in most of our simulations, although in a few cases, e.g.\ in checking the convergence, 
grids of size $1024$, $4096$ and $8192$ are also used. The time step is chosen to be ten times the radial lattice spacing in accordance with the 
smallness of the coordinate light speed.

\subsection{Conserved currents}

As $({\partial}/{\partial T})^\mu$ and $({\partial}/{\partial \varphi})^\nu$ are Killing vectors, the energy and angular momentum currents 
\beq
E^\mu={\mathcal{T}^\mu}_\nu \left(\frac{\partial}{\partial T}\right) ^\nu\ \ \ {\rm and}\ \ \ 
M^\mu={\mathcal{T}^\mu}_\nu \left(\frac{\partial}{\partial \varphi}\right) ^\nu\,
\eeq
are divergence free, where 
\beq
\mathcal{T}_{\mu\nu}=(\nabla_\mu\Phi)(\nabla_\nu\overline{\Phi}) + (\nabla_\nu\Phi)(\nabla_\mu\overline{\Phi}) - g_{\mu\nu}(\nabla_\lambda \Phi)(\nabla^\lambda \overline{\Phi})
\eeq 
is the energy-momentum tensor. It follows then from Stokes' theorem that for a spacetime domain $N$ with boundary $\partial N$ and outward pointing 
unit normal vector $n_\mu$ at $\partial N$, the energy and angular momentum balance relations 
\beq
\label{eq.energy}
\int_{\partial N} n_\mu E^\mu=\int_{N} \nabla_\mu E^\mu =0 \ \ \ {\rm and}\ \ \ \int_{\partial N} n_\mu M^\mu=\int_{N} \nabla_\mu M^\mu =0
\eeq
hold. 
We use these relations---in addition to the usual convergence checks---to verify the correctness and reliability of our code. 
In evaluating the integrals, $N$ is always chosen to be a cylindrical domain, determined by the type of relations $T_1 < T < T_2$, $R_1 < R < R_2$. 
Further details on the calculation of the integrals $\int_{\partial N} n_\mu E^\mu$
and $\int_{\partial N} n_\mu M^\mu$ are given in Appendix B.

\section{Results}
\label{sec.3}

\subsection{Initial conditions}
\label{sec.init}

In our numerical investigations the initial data is always `pure' in the sense that only one of the 
coefficients $\psi_l^m$ or its time derivative is chosen to be non-zero. We consider 
eight different types of initial data labeled by numbers $1,\dots, 8$. The type $1$ and $2$ initial data  are compactly supported whereas for the others 
the physical field $\Phi$ have various fall off property toward future null infinity.

Type $1$ initial data is such that $\Psi=0$ and, for specific values of the indices $(l,m)$, $\partial_T \psi_l^m$ 
is chosen to be $\frac{2R}{1-R^2}$ times the `bump' function 
\begin{align}\label{id1}
\mycal{B}\ &=\ \begin{cases}
	\exp\left({-\frac{1}{|R-c+w/2|}-\frac{1}{|R-c-w/2|}+\frac{4}{w}}\right)\,,
		& \mathrm{if}\ c-w/2\le R  \le c+w/2,\cr
	\hfill 0\,, &\hfill  \mathrm{otherwise}\,,
	\end{cases}
\end{align} 
where $w$ denotes the width of the bump and $c$ determines its center. $\mycal{B}$ is smooth and it is of compact support.
We refer to this initial data as non-stationary as the time derivative of the field is non-zero.

Type $2$ initial data are complementary to type $1$ in the sense that $\psi_l^m$ is chosen to be 
$\frac{2R}{1-R^2}$ times 
the bump function $\mycal{B}$, determined above, while $\partial_T \Psi$ is chosen to be identically zero. 
This type of initial data is going to be referred as stationary since the time derivative of the field vanishes.

For type $1$ and $2$ initial data, in all of our numerical calculations, the values $w$ and $c$ are chosen to be 0.1 and 0.7, respectively, 
i.e.\,the bump is narrow and it is located close but outside the outer event horizon.

The non-compactly supported initial data specifications, type $3$-$8$, also comprise complementary pairs. 
Namely, type $3$, $5$ and $7$ are non-stationary while type $4$, $6$ and $8$ are the corresponding stationary ones yielded 
from $3$, $5$ and $7$ by interchanging the initial data specifications following the rule 
$\Psi_{3,5,7}\rightarrow \partial_T \Psi_{4,6,8}$ and $\partial_T \Psi_{3,5,7}\rightarrow \Psi_{4,6,8}$. 

Type $3$ and $4$ initial data are specified by setting either the coefficient $\psi_l^m$ or its time derivative to be identically one.
In virtue of (\ref{eq.resc}), $\Psi=r\,\Phi$, i.e.\ in this case $\Phi$ or its time derivative 
possesses $1/r$ fall off toward $\scrip$.

In the construction of type $5$-$8$ initial data,  
the bump function $\mycal{B}$ with parameters $c=0.9$ and $w=0.4$ is applied, which vanishes for $R<0.7$, 
i.e.\ near to the outer event horizon, and it is non-zero at $R=1$.  For type $5$ and $6$ either the coefficient $\psi_l^m$ or its time derivative is 
chosen to be $1-R$ times $\mycal{B}$; thereby, it vanishes linearly at $R=1$ 
while it is zero in a neighborhood of the event horizon. Type $7$ and $8$ initial data specifications are constructed analogously multiplying $\mycal{B}$, 
with parameters $c=0.9$ and $w=0.4$, by the factor $(1-R)^2$. 

It is important to note that type $5, 6$ and $7, 8$ initial data specifications imply $1/r^2$ and $1/r^3$ fall off property toward null infinity 
for $\Phi$ or $\partial_T \Phi$, respectively.
In the Schwarzschild case,
initial data with the power law fall off property of type $5, 6$ or $7, 8$ initial data were studied 
by Price \cite{Price}---referred by him as `static' initial data---and Leaver \cite{Leaver}. 
Note also that in \cite{Price,Leaver} the pertinent decay rates were found to be smaller than the corresponding rates 
for initial data with compact support.

\subsection{Power law decay exponents}

After an initial burst phase, the field goes through a longer lasting quasi-normal ringing period while the evolution is closed by a power law decay. 
This, in particular, means that for large values of $T$ the multipole expansion coefficients, at constant $R$,
decrease as $\sim T^{-n}$, where the value of $n=n(l',l)$ is found to be integer 
in analytic investigations \cite{Price, BO, Barack, Hod1, Poisson, Leaver, Kronthaler}. 
Here, using the terminology introduced in \cite{BK,BK3}, the primary multipole index of the pure initial data is referred to as $l'$ while that of the 
examined excited multipole modes as $l$. Since the azimuthal mode 
$m$ is fixed during the evolution, its value will not be explicitly indicated.

The power law decay, in general, starts at $T$ values of a few hundred, where  $T$ is measured in mass units.
In numerical investigations
the decay exponents $n=n(l',l)$ are extracted by 
evaluating the local power index (LPI)
\beq
\frac{d\ln |\psi_l^m|}{d \ln T}\,.
\eeq

In all of our numerical investigations the mass of the black hole $M$ is chosen to be one, while in most of the cases the choice $a=0.5$ is applied, 
although some simulations with $a=0.2$ and $a=0.995$ are also done to justify that the decay rates do not depend on the value of $a$. 
Since in former numerical investigations almost exclusively the $m=0$ case has been considered, 
for the sake of comparison and for simplicity $m=0$ is chosen also in most of our simulations. 
Nevertheless, for type $1$ and $2$ initial data specifications the cases $m=\pm 1$ are also investigated. 

\medskip 

Table \ref{tab1} below shows the decay exponents that we find with type $1$ (i.e.\ compactly supported, non-stationary) initial data. 
\begin{table}[htbp]
\begin{center}
\begin{tabular}{@{}|c||c|c|c|c|c|c|c|c|c|c|}
\hline
$l'$ & \multicolumn{2}{c|}{$l=0$} & \multicolumn{2}{c|}{$l=1$} &  \multicolumn{2}{c|}{$l=2$} & \multicolumn{2}{c|}{$l=3$} & \multicolumn{2}{c|}{$l=4$}   \\
\hline
\hline
0 & 3 & 2 & \multicolumn{2}{c|}{-}  & 5 & 4 & \multicolumn{2}{c|}{-} & 7 & 6  \\
\hline
1 &  \multicolumn{2}{c|}{-} & 5 & 3 & \multicolumn{2}{c|}{-}  & 7 & 5 & \multicolumn{2}{c|}{-}    \\
\hline
2 & 3 & 2 & \multicolumn{2}{c|}{-} &  $\times$  & 4 & \multicolumn{2}{c|}{-} &  $\times$ & 6   \\
\hline
3 & \multicolumn{2}{c|}{-} & 5 & 3 & \multicolumn{2}{c|}{-} &   $\times$ & 5 & \multicolumn{2}{c|}{-}     \\
\hline
4 & 5 & 4 & \multicolumn{2}{c|}{-} & 7 & 4 & \multicolumn{2}{c|}{-} &  $\times$ & 6      \\
\hline
5 & \multicolumn{2}{c|}{-} & 7 & 5 & \multicolumn{2}{c|}{-} &  $\times$ & 5 & \multicolumn{2}{c|}{-} \\
\hline
\end{tabular}
\end{center}
\caption{\label{tab1} Decay exponents $n(l',l)$ for type $1$, non-stationary initial data with compact support, with 
$l=0,1,2,3,4$ and $l'=0,1,2,3,4,5$. Here and elsewhere, in each slot for a specific value of $l$, 
the first number refers to the decay exponent found for $R<1$ while the second entry to that at $R=1$. 
The modes excluded by equatorial reflection symmetry are indicated by the sign "-", while those which remain ambiguous by the sign "$\times$". }
\end{table}
For each value of $l$, two numbers appear in each entry. The first one indicates the decay exponent at constant $R$ 
with $R<1$ (including thereby the outer horizon, $\mathcal{H_+}$) while the second number signifies the decay rate at $R=1$ (representing future null infinity, $\scrip$). For $R<1$, the decay exponents are measured at several values of $R$ 
ranging from the inner boundary of the computational domain to $\scrip$.
The inner boundary of the computational domain is set to $R=0.58$, which is behind the event horizon that is at $R\approx 0.59864$ for $a=0.5$. 
The slots for those exponents that we cannot determine unambiguously---for more details see the related remarks below---are marked by the sign "$\times$". 
The non-stationary and compactly supported type $1$ initial data may be interpreted as data falling off `extremely fast' toward null infinity. 
These type of data were 
used in most of the former numerical simulations and our results for $R<1$ are in agreement with those obtained in \cite{BK,BK3,TKT, GPP, KLP,ZT}. 
In particular, the numbers in the first entries support the rules $n=l'+l+3$ for $l\ge l'$ and $n=l'+l+1$ for $l<l'$ proposed in \cite{BK}.

The numbers in the second entries, indicating the decay rates at future null infinity, are in agreement with those of \cite{ZT}. Nevertheless, 
it is worth mentioning that our results provide extension to that of \cite{ZT} as there only the cases of type 1 initial data with $l'=0\dots 4$ were 
considered  and the decay rates of the entire composite field 
(not the individual multipole components) were determined.
Our results do also fit to the rule $n=l+2$ for $l\ge l'$ and $n=l'$ for $l\le l'-2$, which was obtained analytically in \cite{Hod1}. 
It is interesting, although it is in accordance with this latter rule, that for the excitations $l'=4$ and $l'=5$ two of the lowest allowed excited modes, 
with $l=0,2$ and $l=1,3$, have coinciding decay rates, $n=4$ and $n=5$, at $\scrip$, respectively.

For initial data of type $1$, the cases $m=\pm 1$ are also investigated. For these cases the $l=0$ mode is excluded but 
otherwise---in agreement with \cite{BK}---we find the same decay exponents to apply as listed in Table \ref{tab1}.

As mentioned above, there are certain cases where we cannot determine the decay exponents unambiguously. In the particular cases with $l'=2$, $l=2,4$ and  
with $l'=3$, $l=3$,
the reason
is that the value of $n$ is found to depend on the location, i.e.\ on the particular value of $R$ for $R<1$. In particular, for $l'=2$, $l=2$, 
with $m=0$, the decay exponent is found to be $5$ near the black hole while it takes
the value $7$ for larger $R$
(exceeding the value $4$ characterizing the decay rate at future null infinity). Note that the value $7$ fits to the rule $n=l'+l+3$, whereas $5$ 
corresponds to the rule $n=l'+l+1$.  
In contrast to the $m=0$ case, the decay exponent is found to be $7$ regardless of the value $R<1$ for the cases $m=\pm 1$. 
Analogously, for $l'=2$, $l=4$, we find the exponent $7$ close to the black hole and $9$ far away from it, and for 
$m=\pm 1$ we find that the exponent is $9$ independently of the value of $R<1$. 
For $l'=3$, $l=3$, $m=0$, we find again the exponent $7$ close to the black hole and $9$ far away from it. Interestingly, 
and in contrast to the $l'=2$, $l=2,4$ cases, we find the same type of splitting of the decay rates for $m=\pm 1$. 
This  behavior of the exponents is confirmed by the fact that if the grid size is raised from $2048$ to $4096$, 
then the data improve similarly as shown in Figure \ref{lpi1}. 
These findings definitely deserve further investigations, which, however, are beyond the scope of the present paper as they require 
the use of considerably longer time evolution and, in turn, much higher numerical accuracy.   

\medskip

Table \ref{tab2} shows the decay rates found for type $2$ (stationary and compactly supported) initial data.
\begin{table}[htbp]
%\vspace{2mm}
\begin{center}
\begin{tabular}{@{}|c||c|c|c|c|c|c|c|c|c|c|}
\hline
$l'$ & \multicolumn{2}{c|}{$l=0$} & \multicolumn{2}{c|}{$l=1$} &  \multicolumn{2}{c|}{$l=2$} & \multicolumn{2}{c|}{$l=3$} & \multicolumn{2}{c|}{$l=4$}   \\
\hline
\hline
0 & 3 & 2 &  \multicolumn{2}{c|}{-} & 5 & 4 &  \multicolumn{2}{c|}{-} & 7 & 6  \\
\hline
1 &  \multicolumn{2}{c|}{-}  & 5 & 3 &  \multicolumn{2}{c|}{-}  & 7 & 5 &   \multicolumn{2}{c|}{-}   \\
\hline
2 & 4 & 3 &  \multicolumn{2}{c|}{-} &   $\times$ & 4 &  \multicolumn{2}{c|}{-} &   $\times$ & 6   \\
\hline
3 &  \multicolumn{2}{c|}{-} & 6 & 4 &  \multicolumn{2}{c|}{-} &   $\times$ &  $\times$  &  \multicolumn{2}{c|}{-}     \\
\hline
4 & 6 & 5 &  \multicolumn{2}{c|}{-} & 8 & 5 &  \multicolumn{2}{c|}{-} &   $\times$ &  $\times$       \\
\hline
5 &  \multicolumn{2}{c|}{-} & 8 & 6 &  \multicolumn{2}{c|}{-} &  $\times$  &  $\times$  &  \multicolumn{2}{c|}{-} \\
\hline
\end{tabular}
\end{center}
\caption{\label{tab2}  
Decay exponents $n(l',l)$ for type $2$ compactly supported stationary initial data. 
}
\end{table}

For $l\ge l'$ they are the same as with the first type of initial data, but for $l<l'$ they are greater by one, i.e.\ the rule $n=l'+l+2$ appears 
to hold. Analogous results for Schwarzschild black holes were obtained in \cite{Leaver, KSM, PB, Kronthaler}. The value $n=6$ for $l'=4$, $l=0$ 
was also found in \cite{BK3} in Boyer-Lindquist coordinates. 
It is worth mentioning that for type $2$ initial data the cases of $m=\pm 1$ are also examined and the same exponents are 
found as for the $m=0$ case.
For $l'=2$, $l=2,4$, $m=0$, a similar dependence of the decay exponents on $R<1$ is found as with the 
first type of initial data. For $l'=2$, $l=2$, the exponents $6$ and $7$ are found near the black hole and far from it, 
respectively. For $l'=2$, $l=4$, the exponents $8$ and $9$ are found. For $m=\pm 1$, the exponents,
also similarly as with the first type of initial data, are found to be independent of the value of $R<1$. Their values are 
$7$ and $9$ for $l=2$ and $l=4$, respectively.

\medskip

\begin{figure}[htbp]
\begin{center}
\includegraphics[
  scale=0.75]{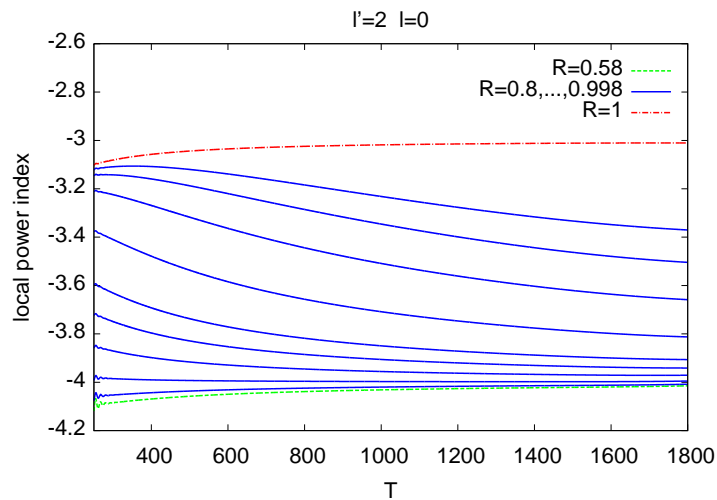}
\caption{\label{lpi2}
{The time dependence of the local power index is shown, for type $2$ initial data with $l'=2$, $l=0$, $m=0$ and $a=0.5$, at  $R=0.58, 0.8, 0.9, 0.95, 0.97, 0.98, 0.99, 0.995, 0.997$, $0.998, 1.0$. The absolute value of the LPI decreases as $R$ increases. The inner boundary of the computational domain is at $R=0.58$, while the event horizon is located at $R\approx 0.59864$.
The LPI at $R=1$ approaches $-3$ as $T$ increases, whereas for $R<1$ the LPI approaches $-4$. 
The grid size applied is $8192$.
}
}
\end{center}
\end{figure}

\begin{figure}[htbp]
\begin{center}
\includegraphics[
  scale=0.75]{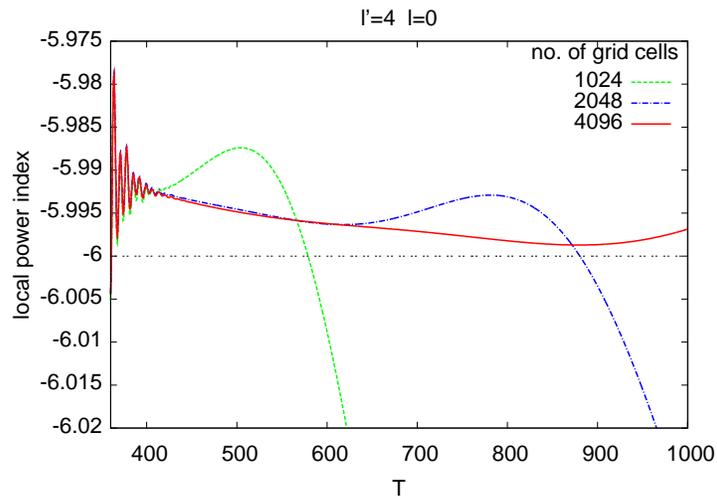}
\caption{\label{lpi1}
{The time dependence of the local power index relevant for grid sizes $1024$, $2048$ and $4096$ is shown 
for type $2$ initial data with $l'=4$, $l=0$, $m=0$ and $a=0.5$ at $R=0.9$,
to illustrate the convergence of the local power index as the number of grid cells together with the time resolution
is increased.
Errors caused by the discretization, resulting here in the dropping down of the LPI for larger $T$,
start to dominate later for larger numbers of grid cells.  
The number of non-zero multipole components taken into account is kept fixed at $8$, corresponding to the cutoff $l_{max}=14$.}
}
\end{center}
\end{figure}

Figure \ref{lpi2} shows the local power index seen at various distances from the black hole for $l'=2$ and $l=0$.
Figure \ref{lpi1} illustrates the convergence of the local power index as the number of grid cells is increased.
At a certain time,
the discretization errors become large in comparison with
the proper amplitude 
of the multipole component, which is a decreasing function of $T$, causing the deviation of the LPI from its correct value.
The error remains small for a longer time if the number of grid cells is increased.   

\medskip

Table \ref{tab3} shows the decay rates found for type $3$ non-stationary initial data. 
\begin{table}[htbp]
%\vspace{2mm}
\begin{center}
\begin{tabular}{@{}|c||c|c|c|c|c|c|c|c|c|c|}
\hline
$l'$ & \multicolumn{2}{c|}{$l=0$} & \multicolumn{2}{c|}{$l=1$} &  \multicolumn{2}{c|}{$l=2$} & \multicolumn{2}{c|}{$l=3$} & \multicolumn{2}{c|}{$l=4$}   \\
\hline
\hline
0 & 2 &  1&  \multicolumn{2}{c|}{-} & 4 & 3 &  \multicolumn{2}{c|}{-} & 6 & 5  \\
\hline
1 &  \multicolumn{2}{c|}{-}  & 4 & 2 &  \multicolumn{2}{c|}{-}  & 6 & 4 &   \multicolumn{2}{c|}{-}   \\
\hline
2 & 2 & 1 &  \multicolumn{2}{c|}{-} &   $\times$ & 3 &  \multicolumn{2}{c|}{-} &   $\times$ & 5   \\
\hline
3 &  \multicolumn{2}{c|}{-} & 4 & 2 &  \multicolumn{2}{c|}{-} &   $\times$ & 4 &  \multicolumn{2}{c|}{-}     \\
\hline
4 & 4 & 3 &  \multicolumn{2}{c|}{-} &  $\times$  &   $\times$  &  \multicolumn{2}{c|}{-} &  $\times$  &   $\times$      \\
\hline
5 &  \multicolumn{2}{c|}{-} & 6 & 4 &  \multicolumn{2}{c|}{-} &  $\times$  &  $\times$  &  \multicolumn{2}{c|}{-} \\
\hline
\end{tabular}
\end{center}
\caption{\label{tab3}
Decay exponents $n(l',l)$ for type $3$ non-stationary initial data with $1/r$ decay for $\Phi$ at $\scrip$. 
}
\end{table}
By comparing these decay exponents to the corresponding ones in Table \ref{tab1}, 
it can be seen that each of the decay exponents found for type $3$ initial data is smaller by one 
than the corresponding exponent for type $1$ initial data. 
In the Schwarzschild case, similar results were found in \cite{Price,Leaver} 
for slowly decaying (they are referred as static in  \cite{Price}) initial data specifications. 
It is also important to note that Dafermos and Rodnianski \cite{DR} in their  recent analytic investigations 
derived $1.5$ as a lower bound on the decay rate of an axial symmetric field in the domain of outer communication, 
$R<1$ in our case, and $0.5$ at $\scrip$, i.e.\ at $R=1$, 
which are just below by one-half with respect to the lowest exponents $2$ and $1$ indicated 
in Table \ref{tab3} for the $l'=0$ and $l'=2$ cases, respectively.

\medskip

Table \ref{tab4} shows the decay rates for type $4$ initial data. 
\begin{table}[htbp]
%\vspace{2mm}
\begin{center}
\begin{tabular}{@{}|c||c|c|c|c|c|c|c|c|c|c|}
\hline
$l'$ & \multicolumn{2}{c|}{$l=0$} & \multicolumn{2}{c|}{$l=1$} &  \multicolumn{2}{c|}{$l=2$} & \multicolumn{2}{c|}{$l=3$} & \multicolumn{2}{c|}{$l=4$}   \\
\hline
\hline
0 & 3 & 2 &  \multicolumn{2}{c|}{-} & 5 & 4 &  \multicolumn{2}{c|}{-} & 7 & 6  \\
\hline
1 &  \multicolumn{2}{c|}{-}  & 4 & 2 &  \multicolumn{2}{c|}{-}  & 6 & 4&   \multicolumn{2}{c|}{-}   \\
\hline
2 & 3 & 2 &  \multicolumn{2}{c|}{-} &  $\times$  & 3 &  \multicolumn{2}{c|}{-} &  $\times$  & 5   \\
\hline
3 &  \multicolumn{2}{c|}{-} & 5 & 3 &  \multicolumn{2}{c|}{-} &  $\times$  & 4 &  \multicolumn{2}{c|}{-}     \\
\hline
4 & 5 & 4  &  \multicolumn{2}{c|}{-} & 7 & 4 &  \multicolumn{2}{c|}{-} &  $\times$  &  $\times$      \\
\hline
5 &  \multicolumn{2}{c|}{-} & 7 & 5 &  \multicolumn{2}{c|}{-} &  $\times$  &  $\times$  &  \multicolumn{2}{c|}{-} \\
\hline
\end{tabular}
\end{center}
\caption{\label{tab4} 
Decay exponents $n(l',l)$ for type $4$ stationary initial data with $1/r$ decay for $\Phi$ at $\scrip$.
}
\end{table}
Notice that the entries are also smaller by one than those for type $2$ initial data, listed in Table \ref{tab2}, with the only exception of the $l'=0$ case when the decay exponents turn out to be identical for type  $2$ and $4$ initial data specifications.

\medskip

Tables \ref{tab5} - \ref{tab8} show the decay rates that we find for type $5$-$8$ initial data. 
\begin{table}[htbp]
%\vspace{2mm}
\begin{center}
\begin{tabular}{@{}|c||c|c|c|c|c|c|c|c|c|c|}
\hline
$l'$ & \multicolumn{2}{c|}{$l=0$} & \multicolumn{2}{c|}{$l=1$} &  \multicolumn{2}{c|}{$l=2$} & \multicolumn{2}{c|}{$l=3$} & \multicolumn{2}{c|}{$l=4$}   \\
\hline
\hline
0 & 3 & 2 &  \multicolumn{2}{c|}{-} & 5 & 3 &  \multicolumn{2}{c|}{-} & 7 & 5  \\
\hline
1 &  \multicolumn{2}{c|}{-}  & 4 & 2 &  \multicolumn{2}{c|}{-}  & 6 & 4 &   \multicolumn{2}{c|}{-}   \\
\hline
2 & 3 & 2 &  \multicolumn{2}{c|}{-} &  $\times$  & 3 &  \multicolumn{2}{c|}{-} &  $\times$   &  $\times$    \\
\hline
3 &  \multicolumn{2}{c|}{-} & 4 & 2 &  \multicolumn{2}{c|}{-} &  $\times$  & 4 &  \multicolumn{2}{c|}{-}     \\
\hline
4 & 5 & 4  &  \multicolumn{2}{c|}{-} & 6 & 3 &  \multicolumn{2}{c|}{-} &  $\times$  &   $\times$      \\
\hline
5 &  \multicolumn{2}{c|}{-} & 6 & 4 &  \multicolumn{2}{c|}{-} &   $\times$ & 4 &  \multicolumn{2}{c|}{-} \\
\hline
\end{tabular}
\end{center}
\caption{\label{tab5} 
Decay exponents $n(l',l)$ for type $5$ non-stationary initial data with $1/r^2$ decay for $\Phi$ at $\scrip$.
}
\end{table}
\begin{table}[htbp]
%\vspace{2mm}
\begin{center}
\begin{tabular}{@{}|c||c|c|c|c|c|c|c|c|c|c|}
\hline
$l'$ & \multicolumn{2}{c|}{$l=0$} & \multicolumn{2}{c|}{$l=1$} &  \multicolumn{2}{c|}{$l=2$} & \multicolumn{2}{c|}{$l=3$} & \multicolumn{2}{c|}{$l=4$}   \\
\hline
\hline
0 & 2 & 1 &  \multicolumn{2}{c|}{-} & 4 & 3 &  \multicolumn{2}{c|}{-} & 6 & 5  \\
\hline
1 &  \multicolumn{2}{c|}{-}  & 4 & 2 &  \multicolumn{2}{c|}{-}  & 6 & 4 &   \multicolumn{2}{c|}{-}   \\
\hline
2 & 4 & 3 &  \multicolumn{2}{c|}{-} &  6  & 3 &  \multicolumn{2}{c|}{-} &  8  &  5   \\
\hline
3 &  \multicolumn{2}{c|}{-} & 5 & 3 &  \multicolumn{2}{c|}{-} &  $\times$  & 4 &  \multicolumn{2}{c|}{-}     \\
\hline
4 & 6 & 5  &  \multicolumn{2}{c|}{-} & 7 & 4 &  \multicolumn{2}{c|}{-} &  $\times$  &  $\times$       \\
\hline
5 &  \multicolumn{2}{c|}{-} & 7 & 5 &  \multicolumn{2}{c|}{-} &   $\times$ & 5 &  \multicolumn{2}{c|}{-} \\
\hline
\end{tabular}
\end{center}
\caption{\label{tab6} 
Decay exponents $n(l',l)$ for type $6$ stationary initial data with $1/r^2$ decay for $\Phi$ at $\scrip$.
}
\end{table}
\begin{table}[htbp]
%\vspace{2mm}
\begin{center}
\begin{tabular}{@{}|c||c|c|c|c|c|c|c|c|c|c|}
\hline
$l'$ & \multicolumn{2}{c|}{$l=0$} & \multicolumn{2}{c|}{$l=1$} &  \multicolumn{2}{c|}{$l=2$} & \multicolumn{2}{c|}{$l=3$} & \multicolumn{2}{c|}{$l=4$}   \\
\hline
\hline
0 & 3 & 2 &  \multicolumn{2}{c|}{-} & 5 & 3 &  \multicolumn{2}{c|}{-} & 7 & 5  \\
\hline
1 &  \multicolumn{2}{c|}{-}  & 5 & 3 &  \multicolumn{2}{c|}{-}  & 7 & 4&   \multicolumn{2}{c|}{-}   \\
\hline
2 & 3 & 2 &  \multicolumn{2}{c|}{-} &  $\times$  & 3 &  \multicolumn{2}{c|}{-} &  $\times$  & 5   \\
\hline
3 &  \multicolumn{2}{c|}{-} & 5 & 3 &  \multicolumn{2}{c|}{-} &  $\times$  & 4 &  \multicolumn{2}{c|}{-}     \\
\hline
4 & 5 & 4  &  \multicolumn{2}{c|}{-} & 6 & 3 &  \multicolumn{2}{c|}{-} &   $\times$ & 5      \\
\hline
5 &  \multicolumn{2}{c|}{-} & 7 & 5 &  \multicolumn{2}{c|}{-} &  $\times$  & 4 &  \multicolumn{2}{c|}{-} \\
\hline
\end{tabular}
\end{center}
\caption{\label{tab7} 
Decay exponents $n(l',l)$ for type $7$ non-stationary initial data with $1/r^3$ decay for $\Phi$ at $\scrip$.
}
\end{table}
\begin{table}[htbp]
%\vspace{2mm}
\begin{center}
\begin{tabular}{@{}|c||c|c|c|c|c|c|c|c|c|c|}
\hline
$l'$ & \multicolumn{2}{c|}{$l=0$} & \multicolumn{2}{c|}{$l=1$} &  \multicolumn{2}{c|}{$l=2$} & \multicolumn{2}{c|}{$l=3$} & \multicolumn{2}{c|}{$l=4$}   \\
\hline
\hline
0 & 3 & 2 &  \multicolumn{2}{c|}{-} & 5 & 4 &  \multicolumn{2}{c|}{-} & 7 & 6  \\
\hline
1 &  \multicolumn{2}{c|}{-}  & 4 & 2 &  \multicolumn{2}{c|}{-}  & 6 & 4&   \multicolumn{2}{c|}{-}   \\
\hline
2 & 4 & 3 &  \multicolumn{2}{c|}{-} & 6 & 3 &  \multicolumn{2}{c|}{-} & 8 & 5   \\
\hline
3 &  \multicolumn{2}{c|}{-} & 6 & 4 &  \multicolumn{2}{c|}{-} & 8 & 4 &  \multicolumn{2}{c|}{-}     \\
\hline
4 & 6 & 5  &  \multicolumn{2}{c|}{-} & 7 & 4 &  \multicolumn{2}{c|}{-} &  $\times$  &   $\times$      \\
\hline
5 &  \multicolumn{2}{c|}{-} & 8 & 6 &  \multicolumn{2}{c|}{-} &   $\times$ & 5 &  \multicolumn{2}{c|}{-} \\
\hline
\end{tabular}
\end{center}
\caption{\label{tab8} 
Decay exponents $n(l',l)$ for type $8$ stationary initial data with $1/r^3$ decay for $\Phi$ at $\scrip$.
}
\end{table}

The following simple observations can be made about these results.
\begin{itemize}
\item[(1)] The decay exponents in Table \ref{tab5} and \ref{tab7} differ by $1$ or $0$ from the exponents listed in Table \ref{tab1}, 
and the same rule applies to Tables \ref{tab6}, \ref{tab8} and \ref{tab2}.

\item[(2)] The decay exponents listed in Table \ref{tab5} are smaller than or equal to those listed in Table \ref{tab7}, 
and the latter are smaller than or equal to those listed in Table \ref{tab1}. 
The same rule applies to Tables \ref{tab6}, \ref{tab8} and \ref{tab2}.

\item[(3)] There are special cases, e.g.\ $l'=4$ for type $5$ initial data, when the dominant, i.e.\ most slowly decaying, 
mode at infinity is definitely not the lowest allowed one.
\end{itemize}

The second observation may be supported by the following argument. If we add  type $1$ and type $5$ initial data, 
then the result is also type $5$, in the sense that it also vanishes as $1-R$ at $R=1$, i.e.\ $\Phi$ falls of as $1/r^2$ at $\scrip$; thus,
it can be expected to yield the decay exponents listed in Table \ref{tab5}. 
On the other hand, due to the linearity of the field equation, 
the late time decay rate yielded by the superposed initial data for a given multipole component is the 
smaller one of the decay rates yielded by the initial data sets that are superimposed. 
Thus, the decay exponents yielded by type $5$ initial data have to be smaller than or equal to the exponents yielded by type $1$ initial data. 
The same reasoning applies to the pairs of type $1$, $7$ and type $5$, $7$ initial data, and to the stationary initial data types as well. 
The assumption on which this argument relies is that the only difference between type $1$, $5$ and $7$ (or $2$, $6$ and $8$) initial data 
that is relevant for the decay rates that they yield is their rate of vanishing at $R=1$.
This argument also implies that initial data specifications for $\Psi$ that tend to zero as $(1-R)^k$, $k\ge 1$---or those which for $\Phi$ have the 
fall off property $1/r^{k+1}$ toward $\scrip$---yield decay exponents that monotonically increase with $k$, 
and are bounded from above by the pertinent decay exponents listed in Table \ref{tab1} or \ref{tab2}.

\subsection{Energy and angular momentum balance}

\begin{figure}[htbp]
\begin{center}
\includegraphics[
  scale=0.75]{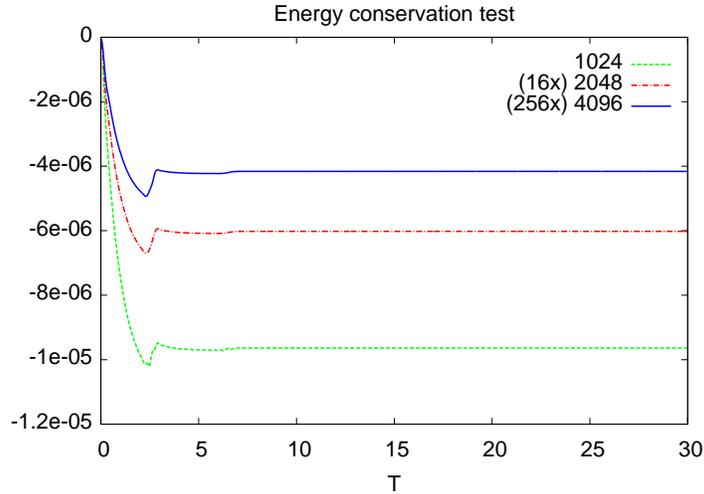}
\caption{\label{f.en}
{The time dependence of the value of $\delta E$, characterizing the numerical violation of 
energy conservation, is shown for type $2$ initial data with $l'=4$ and $m=0$ and for various grid resolutions.
$\delta E$ appears to converge to $0$ at somewhat higher than fourth order rate.
The number of non-zero multipole components taken into account is kept fixed at $8$ as usual, corresponding to the cutoff $l_{max}=14$. 
}
}
\end{center}
\end{figure}

As mentioned above we use the energy balance relation to verify the correctness and reliability of our numerical code. 
Figure \ref{f.en} shows the time dependence of the quantity 
\beq
\delta E=\frac{1}{E_0}\int_{\partial N} n_\mu E^\mu 
\eeq
for various grid sizes. 
Region $N$ is chosen to be the cylindrical domain with $T_1=0$, $T_2=T$, $R_1=R_+$ and $R_2=1$, 
where $R_+$ denotes the location of the outer event horizon, while $E_0$ denotes the initial energy  
\beq
E_0=\int_{T=0,\ R_+\le R \le 1} n_\mu E^\mu\,
\eeq
associated with the part of the $T=0$ hyperboloidal time slice contained by the domain of outer communication.  
In Figure \ref{f.en}, the values of $\delta E$ corresponding to $2048$ and $4096$ grid cells are multiplied by $16$ and $256$, respectively. 
The applied initial data are of type $2$ with $l'=4$ and $m=0$. 
The exact value of $\delta E$ is $0$; thus, its numerically calculated value should be small and tend to zero as the grid size  increases. 
Note that Figure \ref{f.en} 
does also justify that the convergence rate of our numerical simulations is at least four corresponding to the order of the applied 
finite differencing method in the $T - R$ section.

\begin{figure}[htbp]
\begin{center}
\includegraphics[
  scale=0.75]{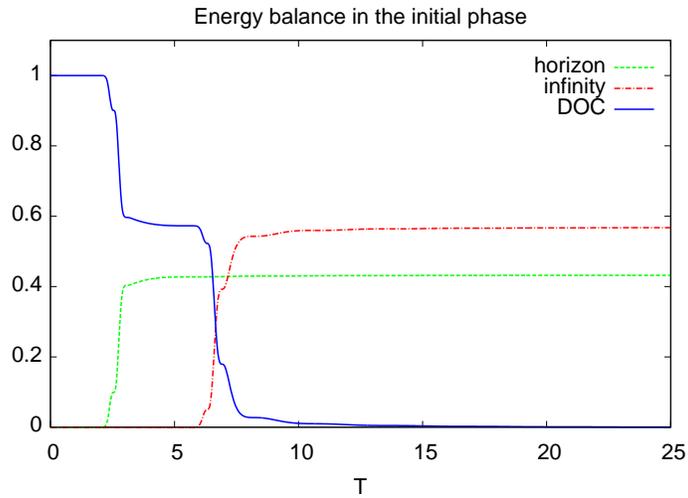}
\caption{\label{en1}
{The time dependence of the normalized total energy in the domain of outer communication (solid blue line), along with the energies flowing 
through the event horizon (dashed green line) and through $\scrip$ (dash-dotted red line) integrated over time, 
during the initial part of the time evolution, for type $2$ initial data with $l'=4$ and $m=0$.
Most of the energy travels in two relatively well-localized and roughly equal-sized bumps, moving toward the event horizon
and toward $\scrip$, respectively.
}
}
\end{center}
\end{figure}

Figure \ref{en1} shows the time dependence of the total energy of the field in the domain of outer communication, 
along with the energies flowing through the event horizon and through $\scrip$,
i.e.\ the integral $\int_0^T n_\mu E^\mu$ of the energy current density at $R_+$ and $R=1$, 
respectively---during the initial phase of the time evolution---for type $2$ initial data with $l'=4$ and $m=0$.
As it is clearly indicated by the graphs the energy of the initial bump splits into two roughly equal sized energy bumps. The inner one reaches the event horizon earlier and a little bit later the outer one leaves the domain of outer communication through $\scrip$. These two bumps carry away almost the entire of the excitation energy.

\begin{figure}[htbp]
\begin{center}
\includegraphics[
  scale=0.75]{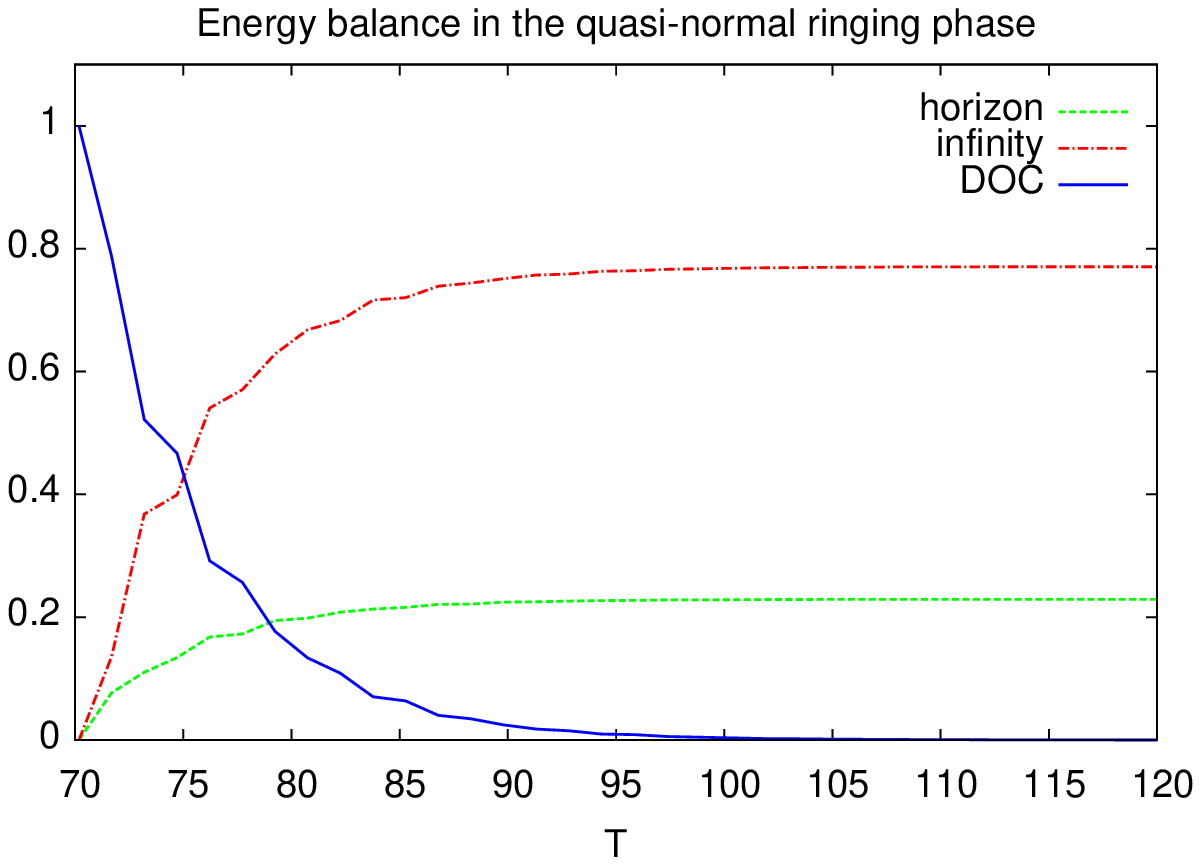}
\caption{\label{en2}
{The time dependence of the normalized total energy in the domain of outer communication (solid blue line), along with the energies flowing through 
the event horizon (dashed green line) and through $\scrip$ (dash-dotted red line) integrated over time from $70$ to $T$, 
is shown for a part of the quasi-normal ringing phase of the evolution for type $2$ initial data with $l'=4$ and $m=0$.
The energy in the domain of outer communication decreases exponentially. 
More energy goes out to infinity than falling into the black hole.  }
}
\end{center}
\end{figure}

Figure \ref{en2} is to show the time dependence of the corresponding quantities during the quasi-normal ringing phase. In this phase considerably larger fraction of the energy is radiated out to $\scrip$ than swallowed by the black hole. 
The total energy of the field in the domain of outer communication decreases as $\sim \exp(-2\,\omega\, T)$, where 
$\omega\approx 0.094$ is the decay constant for the field.  

During the tail phase the  energy flow through the event horizon and through $\scrip$, respectively, 
follows a power law $\sim T^{-\tilde{n}}$ for some integer $\tilde{n}$.
At the event horizon, with $m=0$ and $l'$ even the relation $\tilde{n}=2n+2$ holds, where $n$ denotes 
the decay exponent of the $l=0$ mode of the field, while 
for $l'$ odd $\tilde{n}=2n+1$, where $n$ denotes the decay exponent of the $l=1$ mode. At
$\scrip$, regardless of the parity of $l'$,  the rule
$\tilde{n}=2n+2$ applies, where $n$ denotes the decay exponent (at $\scrip$) of the most slowly decaying mode.  
For $m=\pm 1$, the rule $\tilde{n}=2n+1$ is found to apply at the event horizon 
regardless of the parity of $l'$, where $n$ denotes the decay exponent of the lowest mode, and $\tilde{n}=2n+2$ at
$\scrip$. We should also mention that there are a few cases in which we cannot determine
the value of $\tilde{n}$ properly. 
Since the relevant decay exponents at $\scrip$ are smaller than those at the event horizon, these rules imply that in the tail phase the energy flowing into the black hole becomes, as $T$ increases, negligible in comparison with the energy flowing out to $\scrip$.

Before turning to the description of the angular momentum transports, note that as each term of the contraction $n_\mu M^\mu$, 
appearing in the second relation of (\ref{eq.energy}), is proportional to either $\partial_\varphi \Phi$ or its complex conjugate, 
the study of the angular momentum conservation and transports is interesting only if $m\not=0$.

Regarding the angular momentum conservation, very similar results to those shown in Figure \ref{f.en} are obtained.  
The overall features of the angular momentum transports in the quasi-normal ringing phase and 
in the power law decay phase are also similar to that discussed for the energy transports. 
In particular, during the latter phase, the power law decay $\sim T^{-\tilde{n}}$ of the angular momentum flow with $\tilde{n}=2n$ 
is found to apply at the event horizon and $\tilde{n}=2n+2$ at $\scrip$, where $n$ denotes the decay exponent of the dominant mode of the 
field at the event horizon and at $\scrip$, respectively.

\begin{figure}[htbp]
\begin{center}
\includegraphics[
  scale=0.75]{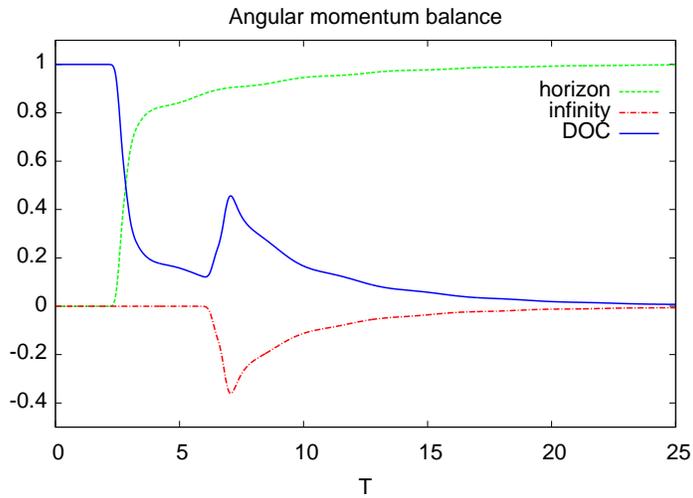}
\caption{\label{ang1}
{The time dependence of the normalized total angular momentum in the domain of outer communication (solid blue line), along with the 
angular momentum flowing through the event horizon (dashed green line) and through $\scrip$ (dash-dotted red line) integrated over time, 
during the initial part of the time evolution, for type $2$ initial data with $l'=4$ and $m=1$.
In contrast with the energy, the angular momentum distribution becomes less localized as time passes. 
Approximately all of the initial angular momentum falls into the black hole in this case.
}
}
\end{center}
\end{figure}

\begin{figure}[htbp]
\begin{center}
\begin{tabular}{cc}
\includegraphics[scale=0.5]{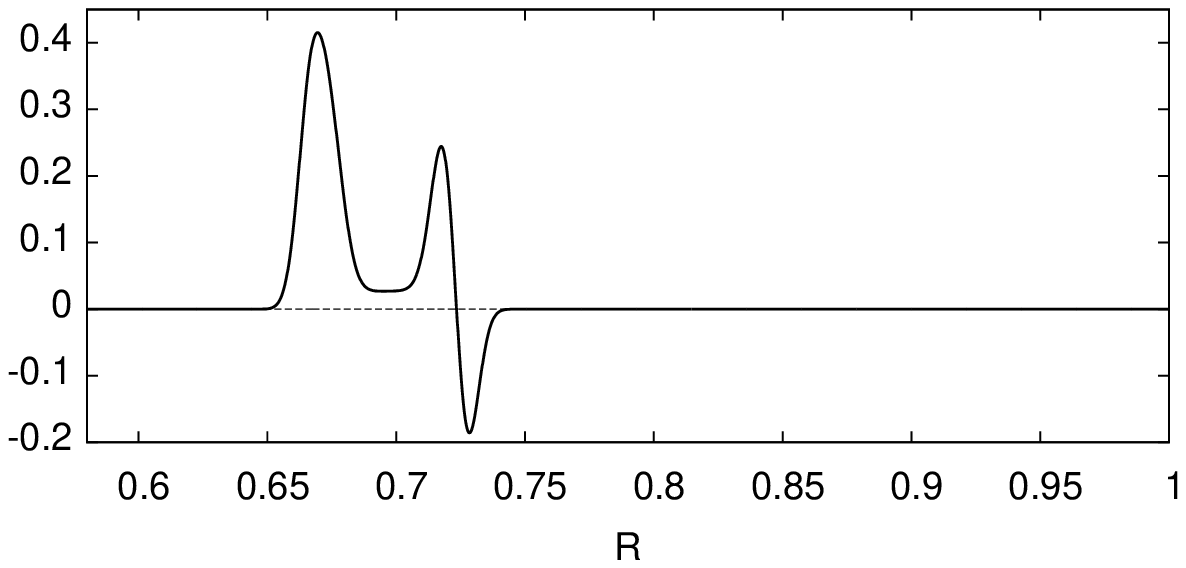} & \includegraphics[scale=0.5]{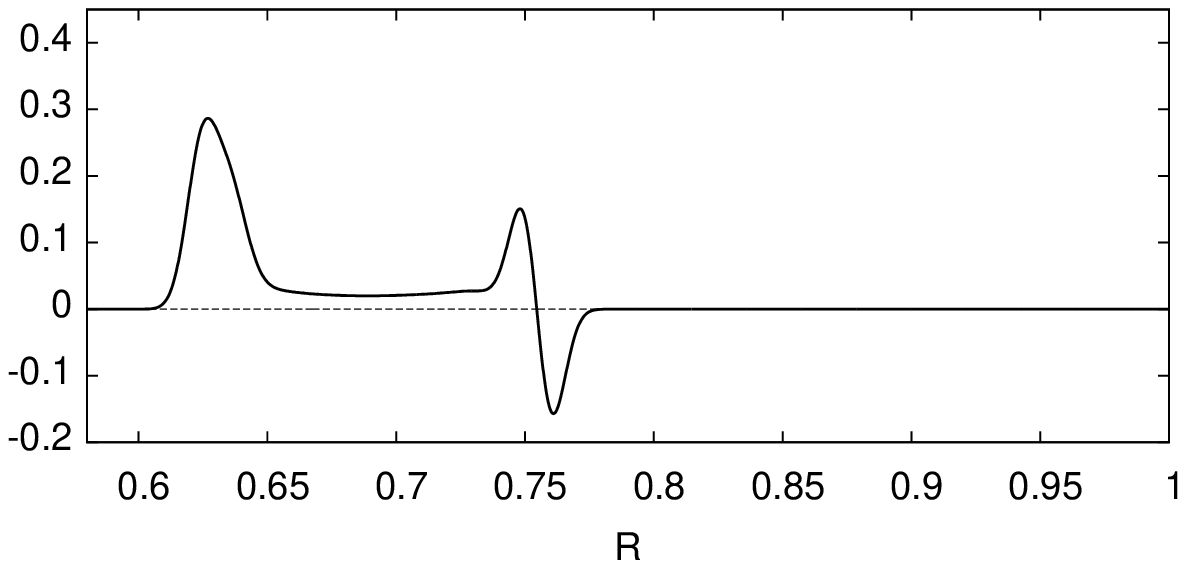}\\[-1.2cm]
\includegraphics[scale=0.5]{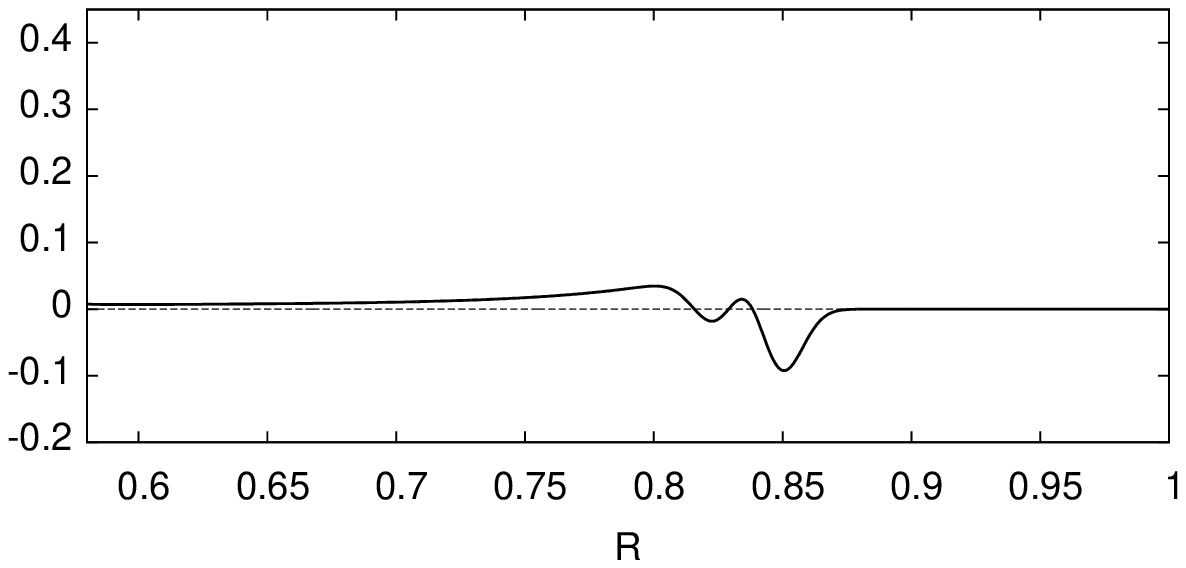} & \includegraphics[scale=0.5]{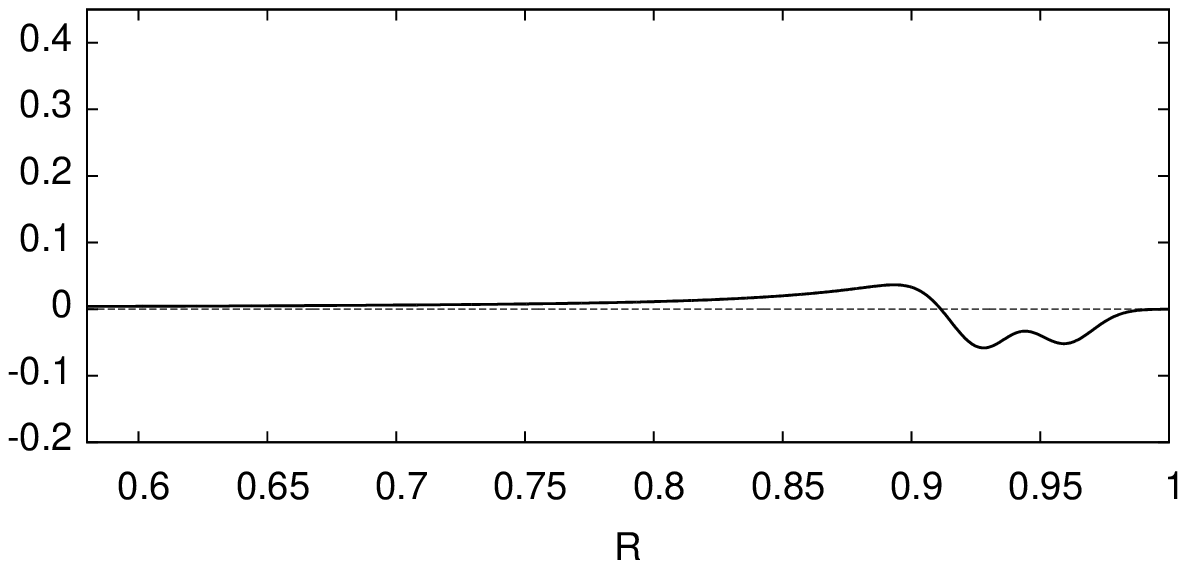}\\
\end{tabular}
\vspace{-0.5cm}
\caption{\label{ang2}
{Angular momentum density as a function of $R$ at $T=1.025$, $T=2.05$, $T=4.1$ and $T=5.84$, for $m=1$, $l'=4$ and type $2$ initial data. 
The outward moving  angular momentum is made up of a positive and a negative part, the sum of which is  approximately zero,
suggesting that the outward moving part of the field is made up of counterrotating shells. 
The density profile of the outgoing angular momentum spreads out in space. 
}
}
\end{center}
\end{figure}

During the initial phase, the time dependence of the angular momentum transfers (see Figure \ref{ang1})
is markedly different from the time dependence of the corresponding energy transports, which are similar to those depicted by Figure \ref{en1}
in the  $m=0$ case.
To make the differences to be more transparent, in
Figure \ref{ang2} the angular momentum density as a function of $R$ for certain time slices is shown.
These graphs indicate 
that almost all of the initial angular momentum moves toward the black hole, while only a small amount moves outward.
The outward moving angular momentum is made up of a positive and a negative part. As time passes, 
the density profile of the outgoing angular momentum spreads out in space, giving rise to the particular time 
dependence of the outflowing angular momentum at $\scrip$ shown in Figure \ref{ang1}.
In spite of the fact that almost all ($97.3\%$) of the energy of the system is lost by $T\sim 8$, in virtue of Figure \ref{ang1} about $30\%$ 
of the initial angular momentum is still present in the domain of outer communication at that time. 

It should be noted that in the considered  case, as indicated by Figure \ref{ang1}, 
the total angular momentum radiated out to infinity tends  approximately to $0$ while $T$ increases. 
However, we find that  
the asymptotic value, to which the angular momentum radiated out toward $\scrip$ tends, depends for instance on the center of the initial bump.

%\newpage

\section{Conclusion}
\label{sec.4}

Our numerical results, concerning the evolution of scalar fields on rotating black hole backgrounds, have been presented. 
The analytic and numerical scheme we have chosen is based on the techniques of conformal compactification and hyperboloidal initial value formulation 
which provided the possibility that the fall off properties of the scalar field along
the event horizon as well as at future null infinity could also be investigated. 

It should be mentioned here that these techniques had already been applied in studying the evolution of linear fields on Kerr background in \cite{ZT}. 
However, our results are more extensive not only because we used a high variety of the initial data specifications, 
but also because the applied $1+(1+2)$ decomposition made the  study of evolution of the individual multipole components possible, 
whereas in \cite{ZT} only the evolution of the entire composite field could be investigated. 
Another technical difference is that while in \cite{ZT} spacelike matching of regions with different properties 
of the applied conformal factors was done---this was admittedly a significant source of numerical error there---we applied 
a single analytic horizon penetrating hyperboloidal foliation which did not require the use of this type of matching.  

\medskip

Our main results can be summarized as follows:

\medskip

In most of the studies of the decay rates of a scalar field on a black hole background investigations are almost exclusively restricted 
to the case of initial data which have extremely fast decay rates toward infinity. To extend the scope of numerical studies we applied, 
besides compactly supported initial data specifications, which clearly belong to the conventional ones, 
other type of initial data specifications possessing $1/r^k$ (with $k=1,2,3$) fall off toward future null infinity. 
In addition, we also studied for each subcases both stationary and non-stationary initial data.
We would like to emphasize, however, that even in the conventional case, all of our results concerning the decay rates 
in the domain of outer communication, including its boundary $\mathcal{H}_+$, are in agreement with those obtained 
in \cite{BK,BK3, TKT, GPP, KLP, ZT}. In particular, the asymptotic power law decay $\sim T^{-n}$ has been confirmed 
with $n=l'+l+3$ for $l\ge l'$ and with $n=l'+l+1$ for $l<l'$ as proposed in \cite{BK}.

Due to the use of conformal compactification, the decay exponents could also be determined at $\scrip$ where our results 
support the analytically obtained predictions of \cite{Hod1} for the power low decay 
with $n=l+2$ for $l\ge l'$ and with $n=l'$ for $l\le l'-2$. In particular, we have found that 
for the excitations $l'=4$ and $l'=5$ two of the lowest allowed excited modes, $l=0,2$ and $l=1,3$, 
have coinciding decay rates, $n=4$ and $n=5$, at $\scrip$, respectively.

We extended the scope of numerical investigations by admitting not only axially symmetric configurations, with azimuthal index $m=0$, 
but for two types of initial data specifications the cases with $m=\pm 1$ were also studied. 
Our findings, on the one hand, confirmed the results of \cite{BK} indicating that 
the decay exponents $n=n(l',l)$ are really independent of the azimuthal index. 
On the other hand, they provided some new insight about the nature of angular momentum transfers.

There were certain situations where the method we applied was not providing definite values for the decay exponents. 
This happened either because the numerical accuracy was not sufficiently high or in some few cases because the 
local power index showed an unexpected location dependence and 'splitting' at the middle 
of the domain of outer communication. 
In particular, we found that in the latter cases the decay rates were smaller in the neighborhood of the black hole 
than in the distant region. We also found that this type of splitting sometimes remains and 
in some other cases disappears if non-axisymmetric
(i.e.\ $m=\pm 1$)
fields are taken into account. These cases definitely deserve further investigations.

In addition to the fall off properties of the scalar field itself, 
we also investigated the energy and angular momentum conservation and transfers. 

Regarding the conservations, our numerical setup was found to be sufficiently accurate and even the convergence rate 
of our numerical code was found, in the initial burst phase, to be larger than four 
in accordance with the order of the applied finite differencing scheme. 

The transports were somewhat more interesting. We found that---in the tail phase---the decay exponents for the 
energy and angular momentum losses at $\scrip$ are smaller than at the horizon which means that 
considerably larger amount of energy and angular momentum is radiated toward future null infinity 
than falling into the black hole. 

We also found that there is significant difference between the behavior of the energy and angular momentum transfers in the initial phase. 
As it is indicated by Figure \ref{ang1}, considerably larger part of the angular momentum of the scalar field 
gets into the black hole than out to $\scrip$. What is even more interesting is that while at the beginning 
the inward moving impulse possesses almost the entire of the initial angular momentum, 
it starts to decrease as the impulse falls toward the black hole, and in the meanwhile, the role of the outward 
moving impulse gets to be more significant. As a result, at the beginning of the quasi-normal ringing phase 
considerably more angular momentum than energy stays in the domain of outer communication.

It would be interesting to investigate the new features of the energy and angular momentum transports in more detail. 
The above-mentioned ambiguities related to some of the fall off exponents would also deserve further investigations, which, however, 
are beyond the scope of the present paper as they require the use of much higher numerical accuracy and considerably longer time evolutions.   
It would also be important to derive analytic predictions for the late time decay exponents that we have found for stationary 
and for slowly falling off initial data.

%\section*{Appendix }
%\label{app.}

\section*{Acknowledgments} 

One of the authors, G.Zs.T., wishes to thank Andr\'as L\'aszl\'o for enlightening discussions related to the applied 1+(1+2) decomposition. 
This research was supported in part by OTKA grant K67942.

%\appendix

\newpage

\section*{Appendix A}
%\label{app.A}

\renewcommand{\theequation}{A.\arabic{equation}} 
\setcounter{equation}{0}

The coefficients appearing in the field equation (\ref{eq.de}) are as follows:

\small

\begin{eqnarray}
a_{RR} & = & \frac{1}{4}(-2a^2 R^2+4R^2-4MR+4MR^3+a^2+a^2 R^4)(-1+R^2)^2 (1+R) \nonumber \\
a_{TR} & = & 2(6MR^4+2a^2R^2+6a^2MR^2+4MR+8M^2R^5+2M-4MR^3-a^2+8M^2R \nonumber   \\
&& -8MR^2-4R^2-16M^2R^3-6a^2MR^4+2a^2MR^6-2a^2M-a^2R^4)(1+R)R \nonumber \\
a_{T\varphi} & = & 4aR(-1+2MR^2-2M)(1+R)(1+R^2) \nonumber \\
a_{R\varphi} & = & a(1+R^2)(1+R)(1-2R^2+R^4) \nonumber \\
a_T & = & \frac{1}{1+R^2}(
a^2+2a^2M-2M+
(a^2+2a^2M-2M)R+
(-4+5a^2+12a^2M-14M)R^2 \nonumber \\
&&+(-48M^2+12a^2M+5a^2-4-30M)R^3+
(-5a^2+4-24a^2M-48M^2+10M)R^4 \nonumber \\
&&+(4-5a^2-24a^2M+32M^2+26M)R^5+
(-a^2+32M^2+4a^2M+6M)R^6 \nonumber \\
&&+(6M+4a^2M+16M^2-a^2)R^7+
(16M^2+6a^2M)R^8+
6a^2MR^9) \nonumber 
\\
a_R & = &\frac{1}{2R(1+R^2)}(
-a^2+
(2M-a^2)R+
(-2a^2+2M)R^2+
(-2a^2+12M)R^3  \nonumber  \\
&& +(12M-12+10a^2)R^4+
(-12+10a^2-24M)R^5+
(8-24M-8a^2)R^6  \nonumber \\
&& +(4M-8a^2+8)R^7+
(4M-a^2+4)R^8+
(-a^2+4+6M)R^9  \nonumber \\
&& +(2a^2+6M)R^{10}+
2a^2R^{11} )
\nonumber \\
a_\varphi & = & -\frac{a(1+R-R^2-R^3)(1+R^2)^2}{R} \nonumber \\
a_0 & = & \frac{1}{2R^2}(
a^2+
(a^2-2M)R
-2MR^2
-2MR^3-
(2a^2+2M)R^4+
(-2a^2+2M)R^5 \nonumber \\
&& +2MR^6+
2MR^7+
(2M+a^2)R^8+
a^2R^9) \nonumber 
\\
a_\Delta & = & (1+R)(1+2R^2+R^4) \nonumber \\
a_{TT}^{(0)} & = & 
4MR+
(32M^2-16a^2M+20M-16M^2a^2+4-4a^2)R^2  \nonumber \\ 
&&+(64M^3-16a^2M+44M+4+96M^2-16M^2a^2-4a^2)R^3  \nonumber \\
&&+(64M^3+32M^2+16a^2M+32M^2a^2-4M)R^4  \nonumber \\
&&+(32M^2a^2-32M^2+16a^2M-64M^3)R^5+
(-64M^3-16M^2a^2)R^6
-16M^2a^2R^7  \nonumber \\
&&+\frac{1}{3}(a^2+a^2R+2a^2R^2+2a^2R^3+a^2R^4+a^2R^5)
 \nonumber \\
a_{TT}^{(2)} & = & \frac{4}{3}\sqrt{\frac{\pi}{5}}(a^2+a^2R+2a^2R^2+2a^2R^3+a^2R^4+a^2R^5)    \nonumber 
\end{eqnarray}

\normalsize

\newpage

\section*{Appendix B}

\renewcommand{\theequation}{B.\arabic{equation}} 
\setcounter{equation}{0}

In this appendix, explicit formulas are given for the integrands in the energy and angular momentum integrals
$\int_{U} n_\mu E^\mu$ and $\int_{U} n_\mu M^\mu$, where $U$ is a $T=const$ or an  $R=const$  hypersurface.
 
On a $T=const$ hypersurface the integral 
$\int_{U} n_\mu E^\mu$ can be written more explicitly as
\beq
\int \intd R\, \intd \theta\, \intd \varphi\   \sqrt{|\det g^{(3)}_{T}|} \ \frac{{\mathcal{T}^T}_T}{\sqrt{|g^{TT}|}},
\eeq 
where $g^{(3)}_{T}$ denotes the metric on the $3$-dimensional $T=const$ surface obtained by the restriction of the full metric.
The integral  $\int_{U} n_\mu M^\mu$ on a $T=const$ hypersurface can be written as 
\beq
\int \intd R\, \intd \theta\, \intd \varphi\   \sqrt{|\det g^{(3)}_{T}|} \ \frac{{\mathcal{T}^T}_\varphi}{\sqrt{|g^{TT}|}}.
\eeq 
Similarly, on an $R=const$ hypersurface, the the energy and  the angular momentum integrals can be written as
\beq
\int \intd T\, \intd \theta\, \intd \varphi\   \sqrt{|\det g^{(3)}_{R}|} \ \frac{{\mathcal{T}^R}_T}{\sqrt{|g^{RR}|}},\qquad
\int \intd T\, \intd \theta\, \intd \varphi\   \sqrt{|\det g^{(3)}_{R}|} \ \frac{{\mathcal{T}^R}_\varphi}{\sqrt{|g^{RR}|}}.
\eeq
Let us introduce the quantities 
\beq
G_1=\frac{1+R}{R},\qquad G_2=-\frac{1+R^2}{R^2}
\eeq
and
%\small
\begin{eqnarray}
c_{TR} & = & 2(2a^2 M R^7 + 8M^2 R^6 + (6M-a^2-6a^2 M)R^5+(-4M-16M^2)R^4 \nonumber\\
&& +(-8M-4+2a^2+6a^2M)R^3+(4M+8M^2)R^2\nonumber\\
&& +(-a^2+2M-2a^2M)R)/(1+R^2)(1+R)^2
\end{eqnarray}

\beq
c_{T\varphi}\ =\ 4(-1+2MR^2-2M)aR/(1+R)^2,\qquad c_{R\varphi}=a
\eeq

\beq
c_{RR}\ =\ (-2a^2R^2-4MR+4MR^3+a^2+4R^2+a^2R^4)/2(1+R^2)
\eeq

\begin{eqnarray}
d_1 & = &  
-2(-16R^7M^2a^2+(-16M^2a^2-64M^3)R^6 \nonumber\\
&& +(-64M^3+32M^2a^2-32M^2+16a^2M)R^5 \nonumber\\
&& +(32M^2a^2+64M^3
           +16a^2M+32M^2-4M)R^4 \nonumber\\
&& +(-16M^2a^2-4a^2+44M-16a^2M+96M^2+4+64M^3)R^3 \nonumber\\
&& +(4-16M^2a^2+20M+32M^2-4a^2-16a^2M)R^2 \nonumber\\
&& +4MR)/(1+R)^3(1+R^2)
\end{eqnarray}

\begin{eqnarray}
d_2 & = &  
-2(a^2R^5+a^2R^4+2a^2R^3+2a^2R^2+a^2R+a^2)/(1+R)^3(1+R^2)
\end{eqnarray}

\beq
c_{TT}\ = \
    (d_1+\frac{d_2}{3})\,2\sqrt{\pi}\, Y_0^0  +  \frac{4}{3}\sqrt{\frac{\pi}{5}}\,d_2 Y_2^0,
\eeq 
%\normalsize
where $Y_0^0$ and $Y_2^0$ are the spherical harmonics
\beq 
Y_0^0=\frac{1}{2}\sqrt{\frac{1}{\pi}},\qquad
Y_2^0=\frac{1}{4}\sqrt{\frac{5}{\pi}}\,(3\cos^2\theta-1).
\eeq

The four integrands above can be expressed as
\begin{eqnarray}
\sqrt{|\det g^{(3)}_{T}|} \ \frac{{\mathcal{T}^T}_T}{\sqrt{|g^{TT}|}} & = & \frac{\sin\theta}{4}\ ( G_1^2 c_{TT}\, \partial_T\Psi\, \partial_T\overline{\Psi} \nonumber \\
&& -c_{RR}(G_2 \Psi +G_1(1-R)\partial_R\Psi)(G_2 \overline{\Psi} +G_1(1-R)\partial_R\overline{\Psi}) \nonumber \\
&& -c_{R\varphi}G_1(1-R)\partial_\varphi\overline{\Psi}(G_2\Psi+G_1(1-R)\partial_R\Psi) \nonumber \\
&& -c_{R\varphi}G_1(1-R)\partial_\varphi\Psi(G_2\overline{\Psi}+G_1(1-R)\partial_R\overline{\Psi}) \nonumber \\
&& +(1+R^2)G_1^2 (\Psi\Delta_{S^2} \overline{\Psi} + \overline{\Psi}\Delta_{S^2} \Psi)/(1+R)^2
)
\end{eqnarray}

\begin{eqnarray}
 \sqrt{|\det g^{(3)}_{T}|} \ \frac{{\mathcal{T}^T}_\varphi}{\sqrt{|g^{TT}|}}
& =  &  \frac{\sin\theta}{4}\ (c_{TT}G_1^2 (\partial_T\overline{\Psi} \partial_\varphi\Psi+\partial_T\Psi \partial_\varphi\overline{\Psi})\nonumber\\
&& +c_{TR}G_1\partial_\varphi\overline{\Psi}(\frac{G_2}{1-R}\Psi+G_1\partial_R\Psi)\nonumber\\
&& +c_{TR}G_1\partial_\varphi\Psi(\frac{G_2}{1-R}\overline{\Psi}+G_1\partial_R\overline{\Psi})\nonumber\\
&& +2c_{T\varphi}G_1^2\partial_\varphi\Psi \partial_\varphi\overline{\Psi} )
\end{eqnarray}

\begin{eqnarray}
 \sqrt{|\det g^{(3)}_{R}|} \ \frac{{\mathcal{T}^R}_T}{\sqrt{|g^{RR}|}}
& = & \frac{\sin\theta}{4}\ ( c_{RR}G_1(1-R)\partial_T\overline{\Psi}(G_2\Psi+G_1(1-R)\partial_R\Psi)\nonumber\\
&& + c_{RR}G_1(1-R)\partial_T\Psi(G_2\overline{\Psi}+G_1(1-R)\partial_R\overline{\Psi})\nonumber\\
&& +2c_{TR}G_1^2\partial_T\Psi\partial_T\overline{\Psi}\nonumber\\
&& +c_{R\varphi}G_1^2(1-R)^2(\partial_\varphi\overline{\Psi}\partial_T\Psi + \partial_\varphi{\Psi}\partial_T\overline{\Psi}))
\end{eqnarray}

\begin{eqnarray}
\sqrt{|\det g^{(3)}_{R}|} \ \frac{{\mathcal{T}^R}_\varphi}{\sqrt{|g^{RR}|}}
& = &  \frac{\sin\theta}{4}\ (c_{RR} G_1 (1-R) \partial_\varphi\overline{\Psi} (G_2\Psi+G_1(1-R)\partial_R\Psi)\nonumber\\
&& + c_{RR}G_1(1-R)\partial_\varphi\Psi(G_2\overline{\Psi}+G_1(1-R)\partial_R\overline{\Psi})\nonumber\\
&& + c_{TR} G_1^2 (\partial_T\overline{\Psi} \partial_\varphi\Psi + \partial_T\Psi \partial_\varphi\overline{\Psi}   )\nonumber\\
&& + 2c_{R\varphi} G_1^2 (1-R)^2 \partial_\varphi\Psi \partial_\varphi\overline{\Psi} )
\end{eqnarray}

The integrals over $\theta\in[0, \pi]$ and 
$\varphi\in[0, 2\pi]$ can be evaluated by considering that
$\int_0^{\pi} \intd \theta \int_0^{2\pi} \intd \varphi\ \sin\theta\ f(\theta,\varphi) = f_0^0$,  where
$f(\theta,\varphi)$ is an arbitrary function and $f_0^0$ is the coefficient of the $l=m=0$ mode in its spherical harmonic expansion. 
The integrals over $T$ and $R$ are evaluated numerically.

\end{document}